\documentclass[12pt]{iopart}
 
\usepackage{graphicx}
\usepackage{xcolor}
\usepackage{float}
\usepackage{url}
\usepackage{hyperref}
\hypersetup{colorlinks   = true,
            citecolor    = blue,
            urlcolor    = blue,
            linkcolor    = black}
\usepackage[USenglish]{babel}
\expandafter\let\csname equation*\endcsname\relax
\expandafter\let\csname endequation*\endcsname\relax
\usepackage{amsmath}
\usepackage[round, sort]{natbib}

\begin{document}

\title[]{Temperature and pressure reconstruction in turbulent Rayleigh-Bénard convection by Lagrangian velocities using PINN}
\author{R. Barta$^{1,2}$, M.-C. Volk$^{1,2}$, C. Bauer$^1$, C. Wagner$^{1,2}$ and M. Mommert$^1$}
\address{$^1$ Institute of Aerodynamics and Flow Technology, German Aerospace Center (DLR), Bunsenstraße 10, Göttingen, 37073, Germany\\
$^2$ Technische Universität Ilmenau, Institute of Thermodynamics and Fluid Mechanics, Ilmenau, Germany}
\ead{robin.barta@dlr.de}
\vspace{10pt}
\begin{indented}
\item[] \textbf{Keywords:} physics informed neural networks, data assimilation, particle tracking velocimetry, Rayleigh-Bénard convection, open-source
\end{indented}
\vspace{5pt}

\begin{abstract}
Velocity, pressure, and temperature are the key variables for understanding thermal convection, and measuring them all is a complex task. In this paper, we demonstrate a method to reconstruct temperature and pressure fields based on given Lagrangian velocity data. A physics-informed neural network (PINN) based on a multilayer perceptron architecture and a periodic sine activation function is used to reconstruct both the temperature and the pressure for two cases of turbulent Rayleigh-Bénard convection (Pr = 6.9, Ra = $10^9$). The first dataset is generated with DNS and it includes Lagrangian velocity data of 150000 tracer particles. The second contains a PTV experiment with the same system parameters in a water-filled cubic cell, and we observed about 50000 active particle tracks per time step with the open-source framework proPTV. A realistic temperature and pressure field could be reconstructed in both cases, which underlines the importance of PINNs also in the context of experimental data. In the case of the DNS, the reconstructed temperature and pressure fields show a 90\% correlation over all particles when directly validated against the ground truth. Thus, the proposed method, in combination with particle tracking velocimetry, is able to provide velocity, temperature, and pressure fields in convective flows even in the hard turbulence regime. The PINN used in this paper is compatible with proPTV and is part of an open source project. It is available on request at \url{https://github.com/DLR-AS-BOA}.
\end{abstract}

\section{Introduction}
\label{intro}
The velocity fields of flows can be measured precisely on temporal and spatial scales, for example, by particle tracking velocimetry (PTV). 
PTV \citep{malik1993particle,maas1993particle,schroder20233d} is a well-known optical measurement technique in experimental fluid mechanics to obtain Lagrangian velocity vector fields within observation domains by tracking the movement of tracer particles within a moving fluid over set time intervals. Many interesting flows in scientific or industrial contexts are thermally driven, such as Rayleigh-Bénard convection (RBC) or the mixed convective flow in the ventilation of closed passenger cabins \citep{schmeling2023numerical}. In both cases, it is much more difficult to measure the temperature field with the same resolution as the velocity fields, which can be achieved, for example, by using thermoliquid crystals as tracer particles for PTV \citep{Kaeufer2023a}. This method is intriguing, but difficult to implement in general settings. Therefore, a more general approach to determine unknown flow properties such as temperature and pressure fields is to reconstruct them by enforcing the underlying governing equations, using known velocity data, typically facilitating physics-informed neural network (PINN) \citep{raissi2019physics}. The PINN approach is rather new, but this topic has increased research activity and interests in recent years \citep{cai2021physics}. Notable examples, relevant for the present application to thermal convective flows, are the reconstruction of velocities from temperatures for synthetic 2D flows \citep{clark2023reconstructing}, the investigation of PINNs framework for full PDE modeling in turbulent convection flows \citep{lucor2022simple} or for background-orient Schlieren measurements \citep{Cai2021}. The opposite reconstruction direction, temperature from velocities, is shown by \cite{mommert2024} for a synthetic (DNS) case of cubic Rayleigh-Bénard convection and by \cite{Toscano2024} for a PTV measurement extended by temperatures from particle image thermometry. \cite{volk2025pinn} further showed for a synthetic case that this methodology can also be applied to the planar datasets of stereoscopic particle image velocimetry. In this paper, we use an extended version of the PINN from \citep{mommert2024,volk2025pinn} based on a multilayer perceptron architecture and a periodic sine activation function to assimilate both temperature and pressure from Lagrangian velocity data generated by direct numerical simulations (DNS) and from a PTV experiment \citep{barta2024large} of turbulent RBC in a cubic cell filled with water. In both cases studied, the Prandtl and Rayleigh numbers are Pr = 6.9 and Ra = $10^9$. \\
RBC \citep{rayleigh1916} is a canonical system for studying thermal convection, in which a fluid is confined in a container and heated from below and cooled from above, ideally with adiabatic sidewalls, and studying RBC is a essential to understanding thermal convection in the big picture. A unified theory that describes the scaling laws of the response parameter (Reynolds and Nusselt number) based on the system parameters in RBC was established by \cite{Grossmann2000}. The set of flow-governing equations is given by the incompressible Navier-Stokes equation in the Oberbeck-Boussinesq approximation and the energy equation for the temperature, both in normalized units.
\begin{align}
\frac{\partial \vec{u}}{\partial t} + (\vec{u} \cdot \nabla)\, \vec{u} &= -\nabla p +\sqrt{\frac{\mathrm{Pr}}{\mathrm{Ra}}}\,\nabla^2 \vec{u} + T \vec{e}_Z,\label{eq:momentum}\\
\frac{\partial T}{\partial t} + \vec{u} \cdot \nabla T &= \sqrt{\frac{1}{\mathrm{Pr}\,\mathrm{Ra}}}\,\nabla^2 T,\label{eq:temperature} \\
\nabla \cdot \vec{u} &= 0.\label{eq:mass}
\end{align}
Here, $\vec{e}_Z$ denotes the unit vector in the vertical direction. Each velocity in equations (\ref{eq:momentum})-(\ref{eq:mass}) has been non-dimensionalized with the free-fall velocity ${u}_{\textit{ref}}=\sqrt{\lambda g \Delta T L}\approx50\,$mm s$^{-1}$, the three spatial coordinates with the cell height ${X}_{\textit{ref}}=L=300$ mm, the time coordinate with the free-fall time $t_{ref}=X_{\textit{ref}}/u_{\textit{ref}}\approx6$ s and the pressure with the reference pressure $p_{\textit{ref}}=\rho u^2_{\textit{ref}}$, where $\rho$ is the fluid density. Also, the pressure is always centered around 0 for comparability. The temperature is always centered around the mean cell temperature and it is non-dimensionalized by $T_{\textit{ref}}= \Delta T$ so that it ranges between -$0.5$ and $0.5$. The PINN minimizes the residual of the three equations (2)-(4) to learn the temperature and pressure fields for a given velocity field which is known at multiple time steps.  \\
The paper is organized as follows. Section \ref{cases} introduces the DNS and experimental Rayleigh-Bénard convection datasets on which the PINN is applied. Section \ref{method} explains the general PINN methodology, including the network architecture and the training routine. In section~\ref{DNS2} the PINN is validated on DNS data and in section \ref{exp2} the temperature and pressure reconstruction with the PINN is tested on experimental data. Finally, in section \ref{concl} a conclusion is drawn.

\section{Rayleigh-Bénard convection datasets}
\label{cases}
In the following subsections, two datasets of turbulent Rayleigh-Bénard convection in a cubic cell (Ra = $10^9$ and Pr = $6.9$) on which the PINN is tested are presented, both containing particle tracks and providing velocities in the Lagrange frame. In the first dataset, the particle tracks are generated in a direct numerical simulation (DNS). The ground truth velocities, temperatures, and pressures are known exactly on the particle positions. The DNS dataset is used to validate the reconstructed temperature and pressure fields with the PINN. The second set contains particle tracks measured with proPTV \citep{barta2023pro} in a cubic cell filled with water \citep{barta2024large}. The chosen reconstruction interval of both datasets is oriented towards the sufficient one of \citep{volk2025pinn}, namely a temporal length of at least $2\, t_{\text{ref}}$.

\subsection{DNS}
\label{DNS}
The DNS is performed in a cubic RBC cell with isothermal heating and cooling plates and adiabatic side walls with Ra $=10^{9}$ and Pr $=6.9$. The DNS solves the dimensionless transport equations for mass, momentum, and temperature for an incompressible fluid using the Oberbeck-Boussinesq approximation, see equations (\ref{eq:momentum})-(\ref{eq:mass}). At all walls, no-slip and impermeability boundary conditions are applied. Furthermore, the dimensionless set of equation (\ref{eq:momentum})-(\ref{eq:mass}) are discretized spatially with fourth-order central differences and in time using a second-order accurate Euler-leapfrog scheme \citep{Wagner1994}. The DNS is performed similarly as discussed in more detail in \citep{barta2023pro} but here for smaller Ra. For the present case of $\mathrm{Pr}>1$, the global mean Batchelor length scale \citep{scheel2013resolving}:
\begin{align}
\label{eq:etaB}
    \eta_B =\eta \, \text{Pr}^{-1/2} = \frac{1}{\text{Ra}^{1/4}\,(\text{Nu}-1)^{1/4}}, 
\end{align}
is smaller than the Kolmogorov length scale $\eta$, and thus more restrictive with respect to the grid resolution. The Batchelor length $\eta_B$ represents the smallest length scale at which temperature fluctuations exist before being dissipated by thermal diffusion. 
Therefore, the minimum grid spacing in the bulk flow region was estimated to resolve the Batchelor scale according to \cite{Groetzbach1983} and \cite{scheel2013resolving}. Furthermore, the minimum grid spacing in the thermal and viscous boundary layers has been estimated according to \cite{Shishkina2010}. For the grid estimation, the Nusselt number can be estimated a priori from the Grossmann-Lohse theory \citep{Grossmann2000,Grossmann2001,Grossmann2002,Grossmann2004}, resulting in $\text{Nu}=61$, which is similar to the Nusselt number calculated a posteriori:
\begin{equation}
	\mathrm{Nu}=\left\langle\sqrt{\text{Ra Pr}}\,T\,\vec{u}\cdot\vec{e}_Z - \frac{\partial T}{\partial Z}\right\rangle_{XYZt}
\label{eq:nusselt}
\end{equation}
where $\langle \cdot \rangle_{XYZt}$ denotes averaging in time, as well as in the $X$, $Y$ and $Z$ directions, resulting in $\mathrm{Nu}=63.4$. Table~\ref{Tab::DNS1} summarizes the simulation parameters, including grid spacing and the number of grid points in the thermal and kinetic boundary layer. 

\begin{table}[H]
\centering
\small  
\setlength{\tabcolsep}{5pt}
\begin{tabular}{|c|c|c|c|c|c|c|c|c|c|}
\hline
Ra & Pr & $N_X\times N_Y \times N_Z$ & $\Delta t$ & $\Delta t_{\text{out}}$ & $\Delta Z_{\text{min}}/L$  & $\Delta Z_{\text{max}}/L$ & $N_{l_T}$ & $N_{l_U}$ & Nu  \\\hline
       $10^{\,9}$ & $6.9$ & $768\times768\times768$ & $10^{\,-4}$ & $0.02$ & $4.5\cdot10^{\,-4}$ & $2\cdot10^{\,-3}$ &  18  &  25 & 63.4    \\
\hline
\end{tabular}
\caption[DNS parameter Ra $=10^{9}$, Pr $=7$]{Ra is the Rayleigh, Pr the Prandtl number. $N_X$, $N_Y$ and $N_Z$ are the number of grid points in $X$, $Y$, and $Z$ direction, respectively. $\Delta t$ is the temporal resolution and $\Delta t_{\text{out}}$ is the temporal output resolution of the particle tracks. $\Delta Z_{\text{min}}$ is the grid spacing at the plates, $\Delta Z_{\text{max}}$ is the grid spacing at the center of the box. $N_{l_T}$ is the number of grid points in the thermal, $N_{l_U}$ in the kinetic boundary layer. The Nusselt number is computed a posteriori.} 
\label{Tab::DNS1}
\end{table}

\begin{figure}[H]
    \centering
    \includegraphics[width=0.8\textwidth]{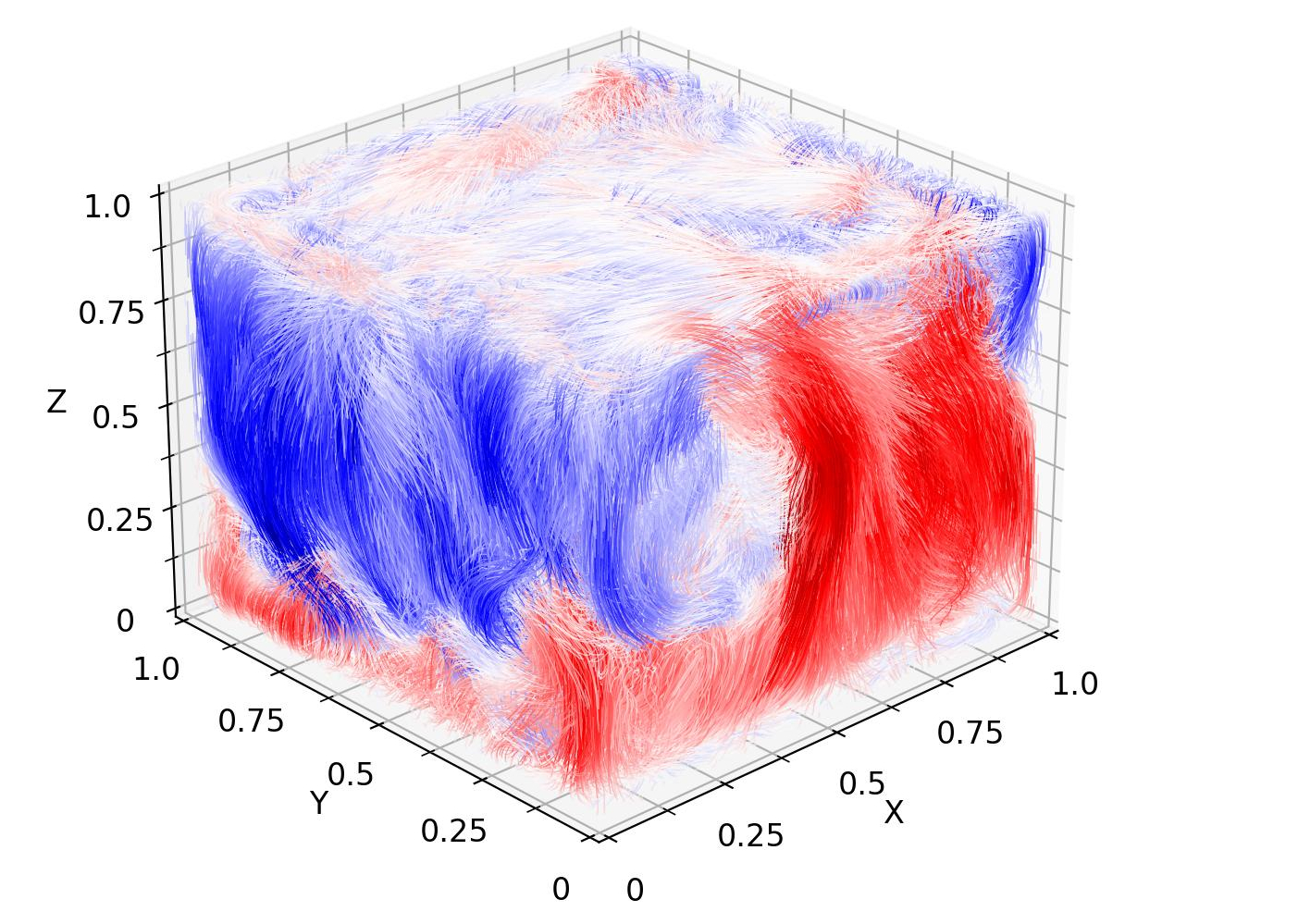} 
    \includegraphics[width=0.16\textwidth]{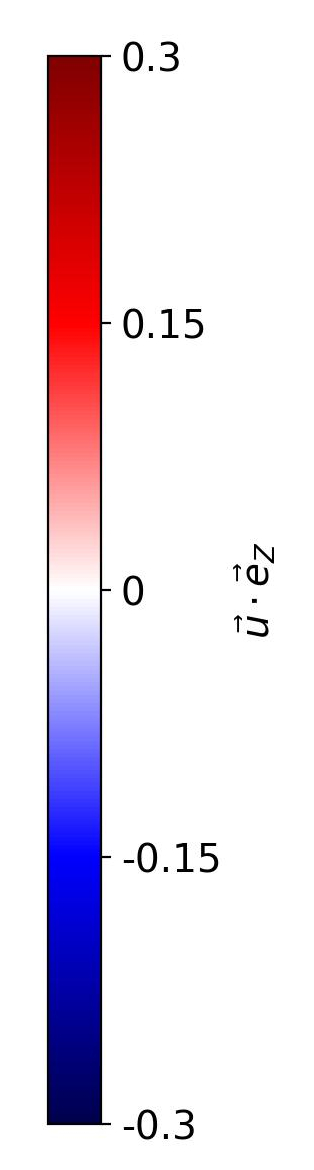}
    \caption{All tracks generated with DNS colored by their vertical velocity.}
    \label{fig:tracksdns}
\end{figure}

After an initial transient, when the average Nusselt number computed reached a statistically stationary state, instantaneous flow field realizations in the form of velocity, pressure, and temperature fields are interpolated on 150000 massless tracer particle positions which are tracked in time. Figure \ref{fig:tracksdns} shows all tracks colored with the vertical velocity. Every $\Delta t_{out} = 0.02$ dimensionless time units, corresponding to a recording frequency of about 8 Hz in the experiment, a particle file is saved for the following analysis, where 125  particle files are used.

\subsection{PTV experiment}
\label{exp}
\begin{figure}[H]
    \centering
    \includegraphics[width=\textwidth]{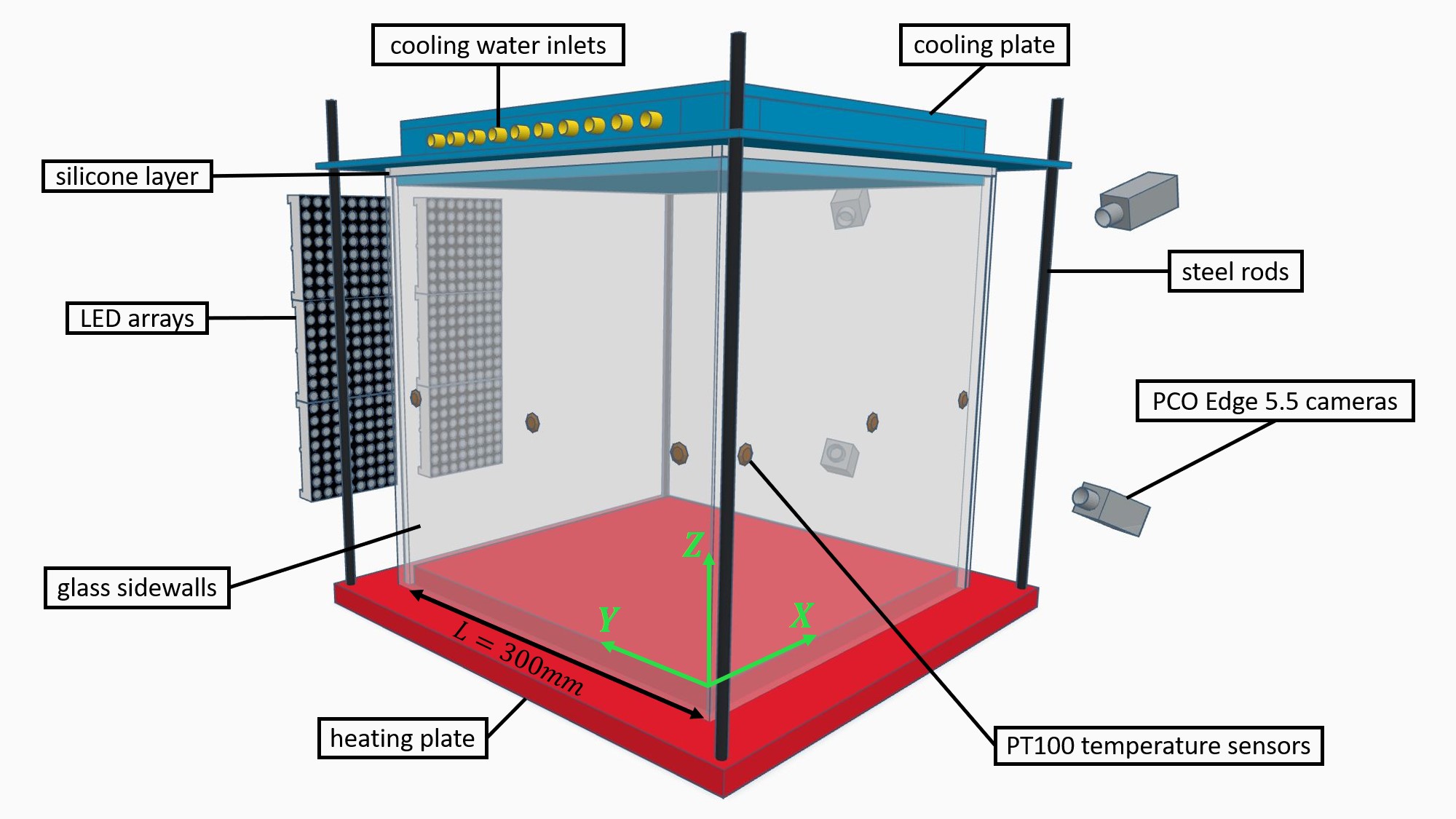} 
    \caption{Technical drawing of the Rayleigh-Bénard experiment. A cubic cell filled with water and heated from below and cooled from above, four cameras observing the flow and two LED arrays illuminating the flow. The green arrows indicate the reference coordinate system.}
    \label{fig:cell}
\end{figure}

The PTV experiment is the same water-filled cubic Rayleigh-Bénard convection cell as discussed in \cite{barta2024large}, except here we use a temperature gradient between the heating and cooling plate of $\Delta T \approx 4\,$K, and a recoding frequency of $f=10\,$Hz. For the temperature difference considered, the Oberbeck-Boussinesq approximation (\ref{eq:momentum})-(\ref{eq:mass}) is valid, as shown in \cite{gray1976validity,weiss2024Ober}. For the experiment, we estimate Ra $\approx10^9$ and Pr $\approx7$ which is similar to the DNS, see Section \ref{DNS}. Figure \ref{fig:cell} shows a technical drawing of the experimental setup and the reference coordinate system. The PTV sequence studied in this work consists of 150 time steps, and image acquisition began after the flow reached equilibrium, approximately 10 hours after seeding tracer particles. In total, 4 PCO Edge 5.5 cameras are observing the flow and each of them has a resolution of 2560$\times$2160 pixels. Polyamide particles manufactured by LaVision with a density of 1.03 g cm$^{-3}$) are used as seeding material. PTV processing is performed using the open source framework proPTV \citep{barta2023pro} and the backtracking routine is used. The image processing detected about 110000 particles per camera. Camera calibration is performed using the built-in Soloff calibration model \citep{soloff1997distortion,herzog2021probabilistic} with a calibration target of 19$\times$19 imprinted markers shifted through the cell at 5 depth positions relative to the cameras, see \cite{barta2024large}. Furthermore, a volumetric correction is applied using particle images such as in \cite{wieneke2008volume} and a subpixel calibration error of less than $0.1$ pixels per camera is achieved. 

\begin{figure}[H]
    \centering
    \includegraphics[width=0.9\textwidth]{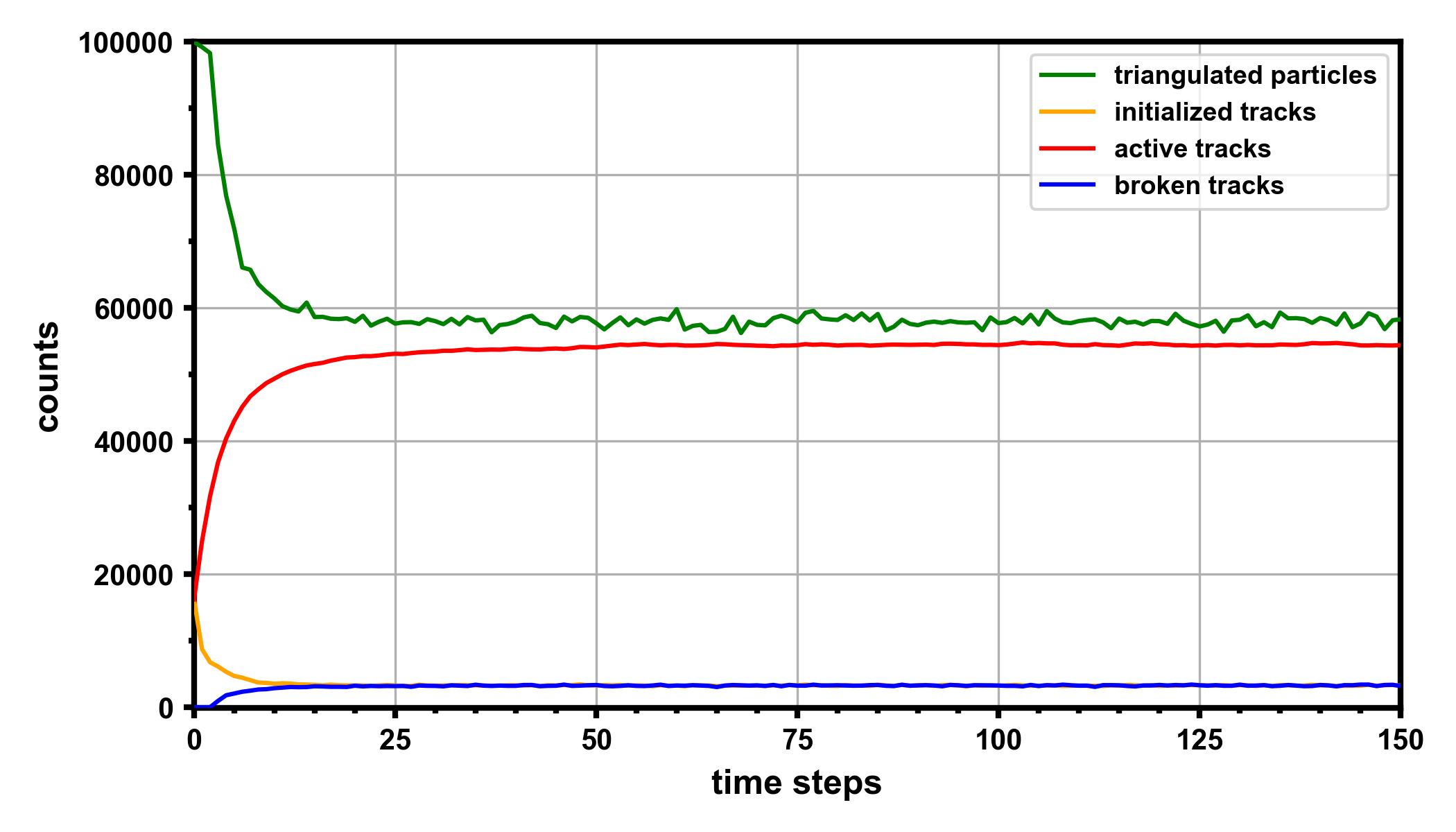}
    \caption{The number of triangulated particles, initialized tracks, active tracks and broken tracks per time step while processing the PTV dataset with proPTV.}
    \label{fig:overview}
\end{figure}

Figure \ref{fig:overview} provides an overview of the number of triangulated particles, initialized tracks, active tracks, and broken tracks during PTV processing of the 150 time steps.

\begin{figure}[H]
    \centering
    \includegraphics[width=0.8\textwidth]{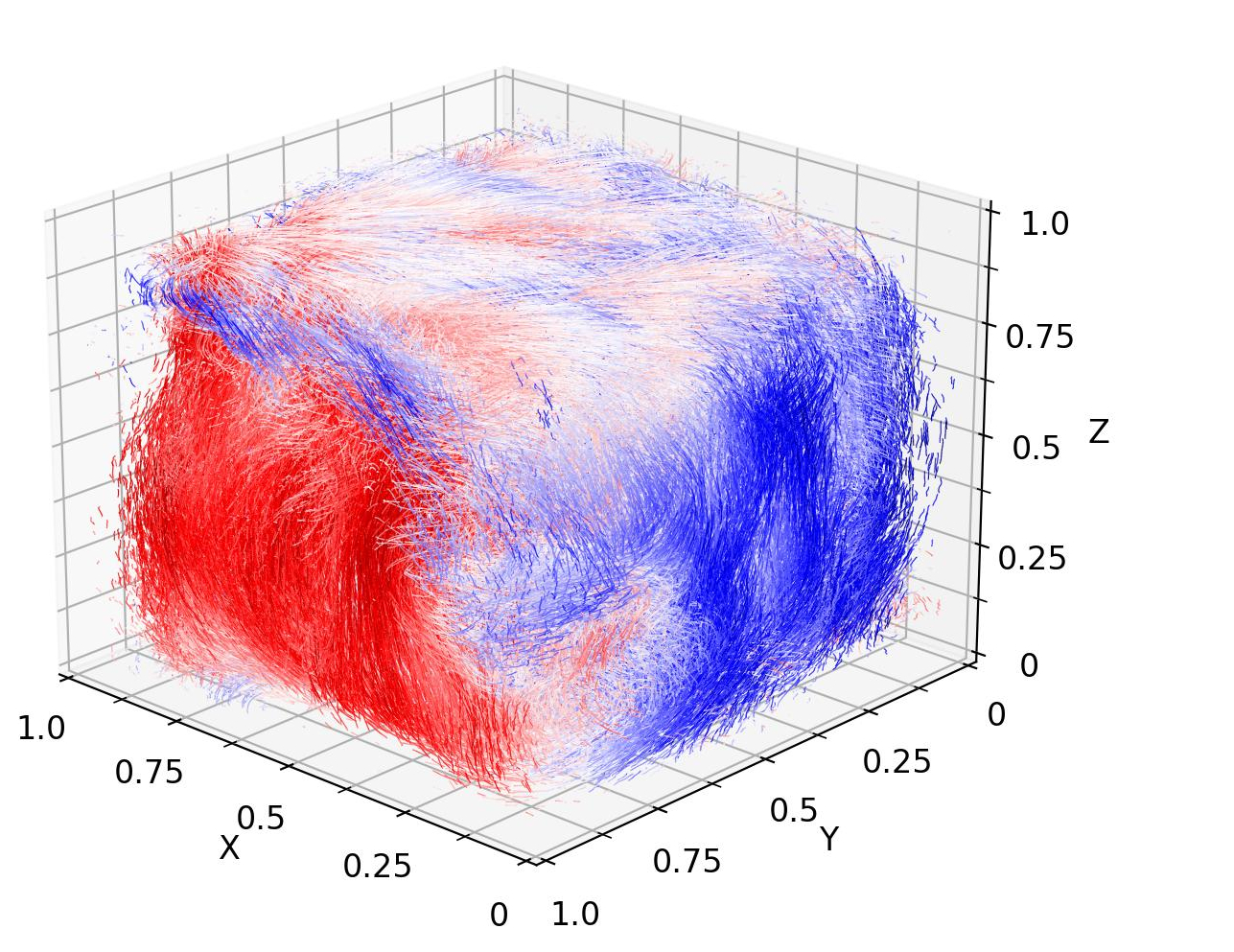} 
    \includegraphics[width=0.145\textwidth]{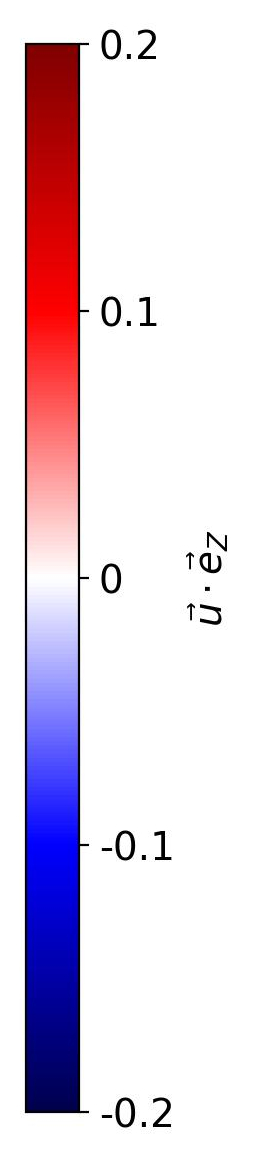}
    \caption{All PTV tracks colored by their normalized vertical velocity.}
    \label{fig:tracks}
\end{figure}

Almost all particles in the camera images are used to triangulate about 100000 particles per time step, from which about 50000 active tracks per time step are reached after processing the 25$^\text{th}$ time step. The new initialized tracks and broken tracks per time step are each about 3000 tracks after the 25$^\text{th}$ time step, so that about 5\% of the tracks break each time step. All tracks obtained by processing the 150 time steps with proPTV are visualized in figure \ref{fig:tracks}, and are colored by their normalized vertical velocity. In units, the maximum observed velocity is about $\pm10$ m s$^{-1}$. During processing, a track is only accepted if it persists for at least 10 time steps. In total about 300000 particle tracks are reconstructed. The track density is high in the bulk flow (about 0.046 ppp), but the track density decreases drastically near the sidewalls because it is observed that the seeding material mostly sticks to the sidewalls when it enters the boundary layer and no reasonable resolution of the flow near the boundaries is achieved. 

\begin{figure}[H]
    \centering
    \includegraphics[width=0.9\textwidth]{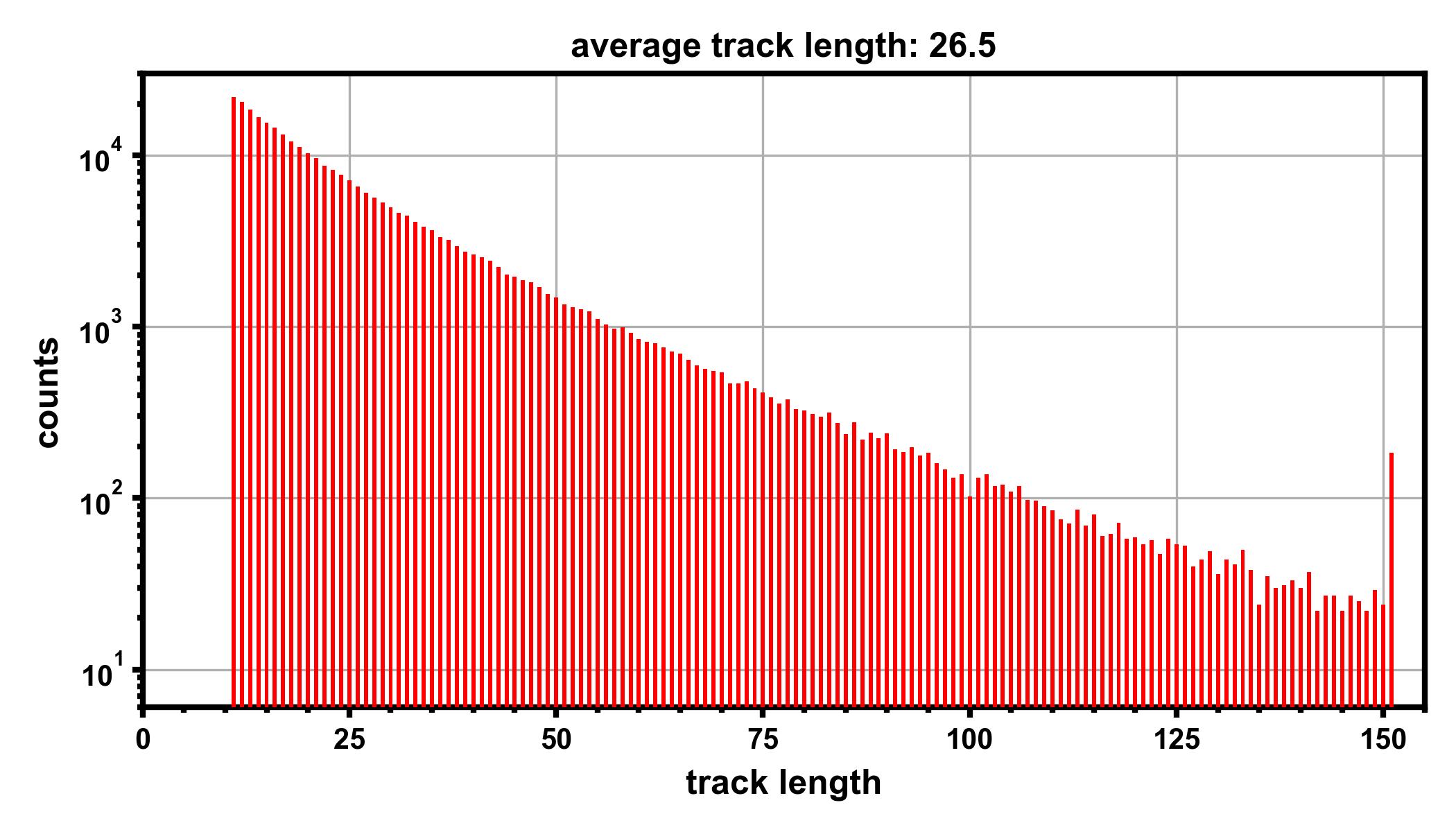}
    \caption{Track length histogram. The average track length is about 26.}
    \label{fig:hist}
\end{figure}

In figure \ref{fig:hist} the track length histogram is shown and an average track length of 26 is calculated. 

In the experiment, the velocity field has full temporal resolution. This is verified because the smallest velocity time scale is larger than the recording time: $\eta/U=0.16\,$s$\,>1/f$, with the mean expected Kolmogorov length in the bulk of $\eta=1.6\,$mm estimated using equation (\ref{eq:etaB}) and the maximum measured velocity magnitude $U\approx10$ m s$^{-1}$. A full spatial resolution of the velocity field is not reached as it would require at least one particle in each volume cell spanned by the Kolmogorov length, or $(L/\eta-1)^3\approx186^3=6435856$ particles. However, the most important structures are the large scale circulation and the corner circulations, which are both integral flow structures and those are well-resolved in our measurement. 

\section{PINN Methodology}
\label{method}
\subsection{Architecture}
\label{CH6:archi}
\begin{figure}[H]
    \centering
        \includegraphics[width=0.7\textwidth]{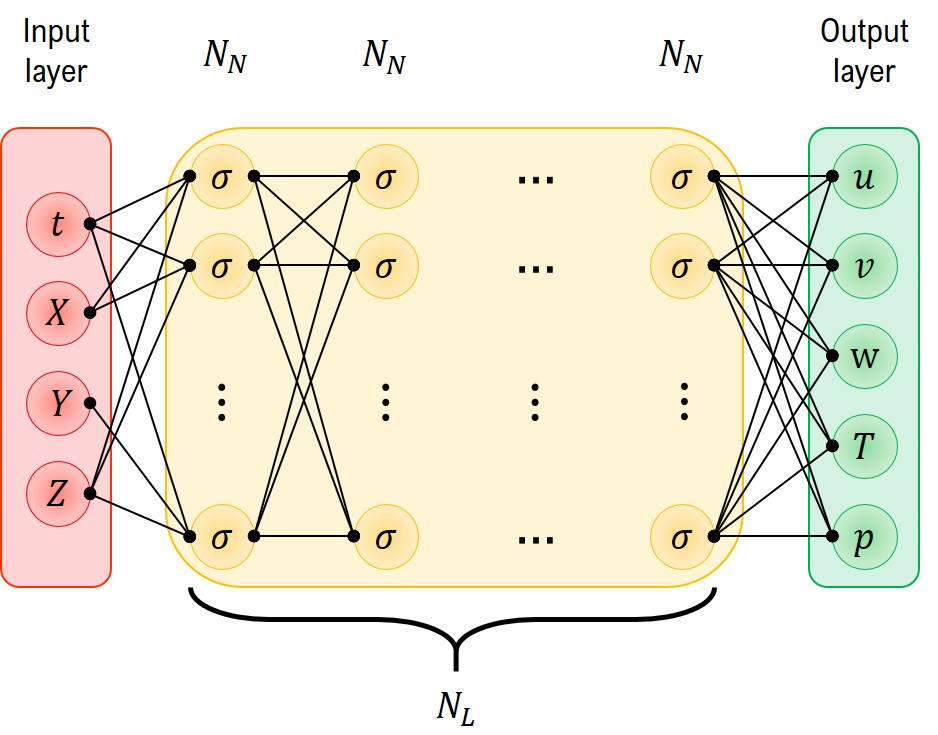}
    \caption[PINN architecture]{Architecture of the MLP used as physics-informed neural network to assimilate temperature and pressure with $N_N$ neurons per hidden layer, $N_L$ hidden layers, and a sinus activation function: $\sigma(\cdot) = \sin(\cdot)$. }
    \label{CH6F::PINNarch}
\end{figure}

A multi-layer perceptron (MLP) is used to reconstruct unknown flow fields, such as temperature and pressure fields from known velocity fields. The MLP, shown in figure \ref{CH6F::PINNarch}, consists of an input layer with 4 neurons corresponding to the time $t$ and position coordinates: $X,Y,Z$, $N_L$ fully-connected hidden layers each with a constant width of $N_N$ neurons, and an output layer with 5 neurons corresponding to the approximated velocity field components $u,v,w$, temperature field $T$ and pressure field $p$. In this paper, we use $N_L=10$ and $N_N=256$. The activation function used is the periodic sine function: $\sigma(\cdot) = \sin(\cdot)$. Sine activation functions are resistant to the vanishing gradient problem \citep{Sitzmann2020}, as their derivative is only close to zero in small periodic regions. The sine activation function helps the MLP to quickly develop an approximation function as a superposition of a large number of wavenumbers, which helps identify multiple structures in turbulent flows. The output layer is linearly activated and does not use an activation function. It should be noted that the proposed PINN architecture can be easily modified to reconstruct only the pressure if necessary. 

\begin{figure}[H]
    \centering
        \includegraphics[width=0.85\textwidth]{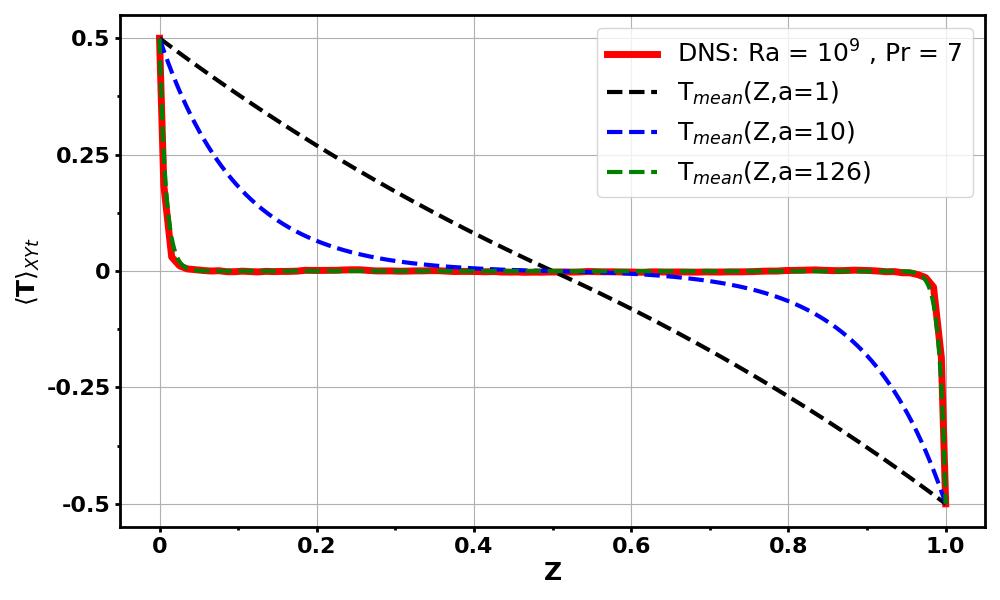}
    \caption[Mean temperature profile]{The mean temperature profile as a function of the vertical cell direction $Z$ averaged with respect to time and the horizontal plane defined by the coordinates $X$ and $Y$ predicted in the DNS for Ra $=10^9$ and Pr $=6.9$. The empirical function $T_{\text{mean}}(Z,a)$ is shown as dashed lines with increasing values of the parameter $a$.}
    \label{CH6F::PINNTmean}
\end{figure}

The network is an extension of the method introduced by \cite{mommert2024}, and it was demonstrated that the PINN can learn the temperature and pressure fields from the provided velocity fields in the cases of RBC in the soft turbulence regime with low to medium Rayleigh numbers \citep{mommert2024,volk2025pinn}. That the simultaneous temperature and pressure reconstruction performs well in such cases is still remarkable because the model solves an ambiguity in the fundamental equations, since both temperature and pressure are unknown and both variables are in the same equation (\ref{eq:momentum}). The context of the given velocity data and the moderately complex loss landscape of flows in the soft turbulence regime are sufficient to resolve this ambiguity in the equations. However, it becomes problematic in RBC at high Rayleigh numbers, particularly in the hard turbulence regime $\text{Ra}>4\cdot10^7$ \citep{castaing1989scaling}, since then the loss landscape is highly complex, with many small-scale structures in the flow. Additionally, the high temperature gradients near the heating and cooling walls such as the near zero temperature in the well-mixed bulk are numerically difficult to resolve, and the reconstruction of both pressure and temperature are extremely challenging for a PINN. We address this problem by introducing a mean temperature profile to the PINN, and only the correct temperature fluctuations in addition to the pressure need to be learned. Thus, the temperature output of the PINN is modified by:

\begin{align}
\label{eq:TmeanPINN}
    &T = T_{\text{mean}}(Z,a) + \left(Z^2-Z\right)\,T' \\
    &\quad \quad \text{with} \quad T_{\text{mean}}(Z,a) = \frac{1}{2-2\,\text{e}^{-0.5\,a}}\,
    \begin{cases}
    \text{e}^{-a\,Z}-\text{e}^{-0.5\,a} & \text{, if } Z < 0.5 \\
    \text{e}^{-0.5\,a}-\text{e}^{\,a\,(Z-1)} & \text{, if } Z \geq 0.5
    \end{cases}
\end{align}

representing a mean temperature profile along the vertical $Z$ axis, $T'$ represents the temperature fluctuation field, and $a$ is a user-defined parameter. A similar approach was used by \cite{Toscano2024}, but they proposed to prescribe a linear mean temperature profile which turned out to be not helpful in the present case at Ra $=10^9$. Equation (\ref{eq:TmeanPINN}) is chosen empirically and ensures that the temperature boundary conditions are satisfied exactly.  The physical meaning of $a$ is derived by evaluating equation (\ref{eq:nusselt}) on the heating or cooling plate where the velocity is zero, and using the derivative of equation (\ref{eq:TmeanPINN}) with respect to $Z$. Thus, we can relate the parameter $a$ with the Nusselt number:
\begin{align}
    \text{Nu} = \frac{a}{2-2\,\text{e}^{-0.5\,a}} \simeq \frac{a}{2}\,, \quad \text{if} \quad a\gg1\,,
\end{align}
and set $a=2\,\text{Nu}=126$ throughout the paper, because we know that the average Nusselt number is Nu $=63$ obtained from DNS in our case, see section \ref{DNS}. Figure \ref{CH6F::PINNTmean} illustrates how the mean temperature profile evolves as the Nusselt number in terms of the parameter $a$ increases. Note that the parameter value $a=1$ results in a linear temperature profile, similar to the one used by \cite{Toscano2024}. 

\subsection{Training and optimization}
\label{CH6:training}
The purpose of the PINN is to reconstruct the temperature and pressure fields based on velocity fields obtained from experiments or flow simulations using the flow-governing equations of RBC in dimensionless units (\ref{eq:momentum})-(\ref{eq:mass}). The training of the PINN over $N_{\text{epoch}}$ epochs is implemented in a Python framework using Tensorflow v.2.16 and Keras v.3.4 \citep{Abadi2016,Chollet2015} with the MLP architecture described in section \ref{CH6:archi}. Specifically, the Adam optimizer \citep{Kingma2014} is used in the default configuration, except for a learning rate schedule that starts with a learning rate, usually $\text{lr}_{\max}=10^{-3}$, and reduces it by a factor of 0.8 until it reaches a certain value, usually $\text{lr}_{\min}=10^{-4}$. The learning rate reduction is applied after $N_{\text{lr}}$ epochs of training have passed. A constant batch size of $N_B=4096$ data points is used, which is a small fraction of the dataset studied. A smaller batch size means more optimization runs per epoch, which accelerates convergence \citep{Sankaran2022} at the cost of computing time. The optimization of the internal PINN parameters is done by optimizing the following total loss function.

\begin{align}
\label{Eq:totL}
    \mathcal{L}_{\text{tot}} = \lambda_{\text{data}}\, \mathcal{L}_{\text{data}} + \lambda_{\text{NS}}\, \mathcal{L}_{\text{NS}} + \lambda_{\text{EE}}\, \mathcal{L}_{\text{EE}} + \lambda_{\text{div}}\, \mathcal{L}_{\text{div}} + \lambda_{\text{C}}\, \mathcal{L}_{\text{C}} + \lambda_{\text{BC}}\, \mathcal{L}_{\text{BC}}
\end{align}

\begin{align}
\label{eq:lossesterms}
    &\mathcal{L}_{\text{data}} = \frac{1}{3\,N_B} \sum_{i=1}^{N_B}\,\left\Vert \left(\vec{u}_i-\vec{u}^*_i\right)^2 \right\Vert  \\
    &\mathcal{L}_{\text{NS}} = \frac{1}{N_B} \sum_{i=1}^{N_B^{\,\text{col}}}\,\left\Vert \left( \frac{\partial\,\vec{u}_i}{\partial\,t} + \left(\vec{u}_i\cdot\nabla\right)\vec{u}_i - \sqrt{\frac{\text{Pr}}{\text{Ra}}}\,\Delta\,\vec{u}_i + \nabla\,p_i - T_i\,\vec{e}_Z\right)^2 \right\Vert \\
    &\mathcal{L}_{\text{EE}} = \frac{1}{N_B} \sum_{i=1}^{N_B^{\,\text{col}}}\,\left\Vert \left( \frac{\partial\,T_i}{\partial\,t} + \left(\vec{u}_i\cdot\nabla\right)T_i - \sqrt{\frac{1}{\text{Pr Ra}}}\,\Delta\,T_i \right)^2 \right\Vert \\
    &\mathcal{L}_{\text{div}} = \frac{1}{N_B} \sum_{i=1}^{N_B^{\,\text{col}}}\,\left\Vert \left( \nabla\cdot\vec{u}_i \right)^2 \right\Vert \\
    &\mathcal{L}_{\text{C}} = \frac{1}{N_B} \sum_{i=1}^{N_B}\,\left\Vert \left( p_i \right)^2 \right\Vert \\
    &\mathcal{L}_{\text{BC}} = \frac{1}{N_{\text{B}}} \sum_{i=1}^{N_{\text{B}}^{\,\text{col}_X}} \, \left\Vert \vec{u}(t_i, 0, Y_i, Z_i)^2 \right\Vert + \left\Vert \vec{u}(t_i, 1, Y_i, Z_i)^2 \right\Vert + \\
    &\quad\quad\quad\frac{1}{N_{\text{B}}} \sum_{i=1}^{N_{\text{B}}^{\,\text{col}_Y}} \, \left\Vert \vec{u}(t_i, X_i, 0, Z_i)^2 \right\Vert + \left\Vert \vec{u}(t_i, X_i, 1, Z_i)^2 \right\Vert + \notag\\
    &\quad\quad\quad\frac{1}{N_{\text{B}}} \sum_{i=1}^{N_{\text{B}}^{\,\text{col}_Z}} \, \left\Vert \vec{u}(t_i, X_i, Y_i, 0)^2 \right\Vert + \left\Vert \vec{u}(t_i, X_i, Y_i, 1)^2 \right\Vert \notag
\end{align}

The total loss function used to train the PINN is a sum of several mean squared error loss contributions with their corresponding global weight factors $\lambda$. The loss contributions are: the data loss $\mathcal{L}_{\text{data}}$, the pressure centering loss $\mathcal{L}_{\text{C}}$ which restricts the pressure to be centered around zero, the boundary loss $\mathcal{L}_{\text{BC}}$ for the velocity, and physical losses: the residual loss $\mathcal{L}_{\text{NS}}$ of the momentum equations (\ref{eq:momentum}), the residual loss $\mathcal{L}_{\text{EE}}$ of the energy equation (\ref{eq:temperature}), and the loss $\mathcal{L}_{\text{div}}$ due to the incompressibility condition (\ref{eq:mass}). The data given in the data loss term equation (\ref{eq:lossesterms}), e.g. from a DNS or a measurement, are denoted by an asterisk, e.g. $\vec{u}^*$ for a known velocity field. The physical loss terms $\mathcal{L}_{\text{NS}}$, $\mathcal{L}_{\text{EE}}$ and $\mathcal{L}_{\text{div}}$ are not evaluated at the same data points given by the batch $N_B$ as the data loss $\mathcal{L}_{\text{data}}$ and the pressure centering loss $\mathcal{L}_{\text{C}}$ during training. Instead, a new collocated batch of the same size as $N_B$, called $N_B^{\text{col}}$, is defined per training epoch, containing input data points $(t_i,X_i,Y_i,Z_i)$ randomly uniform distributed within the cubic cell and across all data time steps. 

Thus, assuming a long enough training time, the physical loss terms are estimated nearly everywhere in the domain, providing the spatial resolution necessary to capture flow structures on a wide range of scales \citep{hou2023enhancing}. Furthermore, three other collocation batches $N_{\text{B}}^{\,\text{col}_X}$, $N_{\text{B}}^{\,\text{col}_Y}$, and $N_{\text{B}}^{\,\text{col}_Z}$ are defined. Each of them contains $N_B$ randomly uniform distributed input data points, in each of the two pairs of opposite sidewalls, e.g. $X=0$ and $X=1$ in the batch $N_{\text{B}}^{\,\text{col}_X}$ and similarly for the batches with superscript $Y$ and $Z$. The three collocation batches on the sidewalls of the cube are used to train the loss of the boundary condition $\mathcal{L}_{\text{BC}}$ that suppresses the velocity to zero at each sidewall. In the following, we use fixed weight factors associated with the values shown in table \ref{Tab::GlobW}, which are chosen following the arguments given by \cite{mommert2024}.

\begin{table}[H]
\centering
\begin{tabular}{|c|c|c|c|c|c|}
\hline
$\lambda_{\text{data}}$ & $\lambda_{\text{NS}}$ & $\lambda_{\text{EE}}$ & $\lambda_{\text{div}}$ & $\lambda_{\text{C}}$ & $\lambda_{\text{BC}}$\\
\hline
1.0 & 0.1 & 0.01 & 0.001 & 0.001 & 0.0001 \\
\hline
\end{tabular}
\caption{Global weights used in the total loss function of the PINN.}
\label{Tab::GlobW}
\end{table}

\subsection{Evaluation metrics}
Two metrics are used to monitor the PINN training process: the mean average error MAE$_{\zeta}$ and the Pearson correlation coefficient PCC$_{\zeta}$ with $\zeta\in\left\{u,v,w,T,p\right\}$, and they are defined by:

\begin{align}
    \text{MAE}_{\zeta} = \frac{1}{N_s}\sum_{i=1}^{N_s}\,\left\Vert \zeta_i-\zeta_i^* \right\Vert \,,
\end{align}
\begin{align}
    \text{PCC}_{\zeta} = \frac{N_{\text{s}} \sum_{i=1}^{N_{\text{s}}} \zeta_i \zeta^*_i - \sum_{i=1}^{N_{\text{s}}} \zeta_i \sum_{i=1}^{N_{\text{s}}} \zeta^*_i}{\sqrt{N_{\text{s}} \sum_{i=1}^{N_{\text{s}}} \zeta_i^2 - \left(\sum_{i=1}^{N_{\text{s}}} \zeta_i\right)^2} \sqrt{N_{\text{s}} \sum_{i=1}^{N_{\text{s}}} (\zeta^*_i)^2 - \left(\sum_{i=1}^{N_{\text{s}}} \zeta^*_i\right)^2}}\,.
\end{align}
Here, $N_s$ denotes the number of samples used to calculate the individual metrics. Typically, the last 10 time steps of each dataset are used to evaluate the metrics, according to \cite{volk2025pinn}, who found that the performance of the PINN is best evaluated after about 2 free fall units in our case. The terms denoted with an asterisk correspond to given data from a DNS or a measurement.


\section{PINN validation using synthetic PTV data}
\label{DNS2}
The results obtained with the PINN are validated in comparison with results obtained from 150000 synthetic particle trajectories generated in a DNS (section \ref{DNS}) of a hard turbulent RBC with Ra $=10^{9}$, Pr $=6.9$, which was processed for 125 time steps representing 2.5 free fall time units. The parameter $a=126$ is chosen to mimic the mean temperature profile and the temperature boundary conditions at the top and bottom plates for the considered case with a Nusselt number Nu $= 63$. The PINN was trained over 2500 epochs and the training process took about 55 seconds per epoch on a NVIDIA GeForce RTX 4090. 

\begin{figure}[H]
    \centering
        \includegraphics[width=0.49\textwidth]{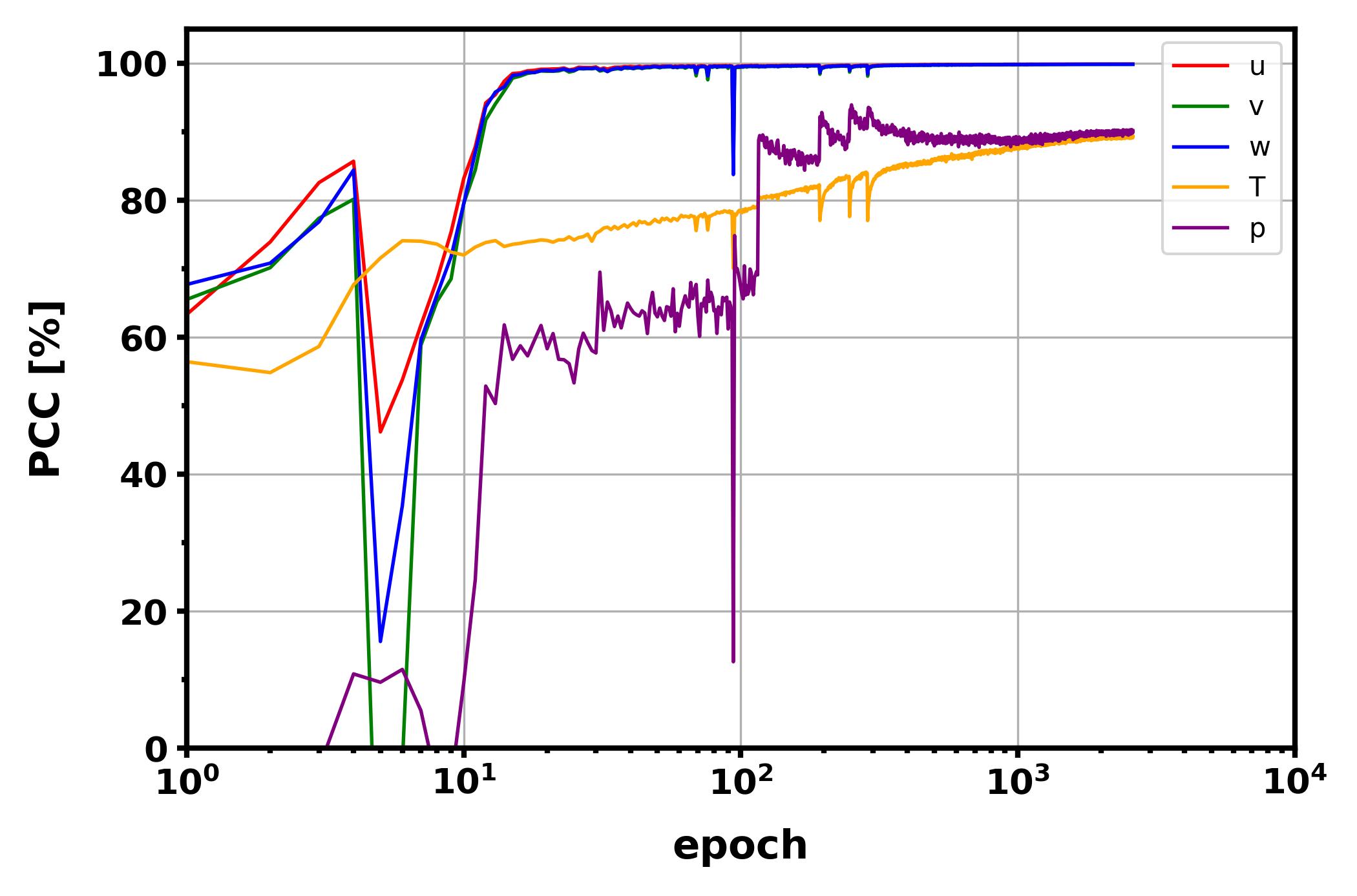}
        \includegraphics[width=0.49\textwidth]{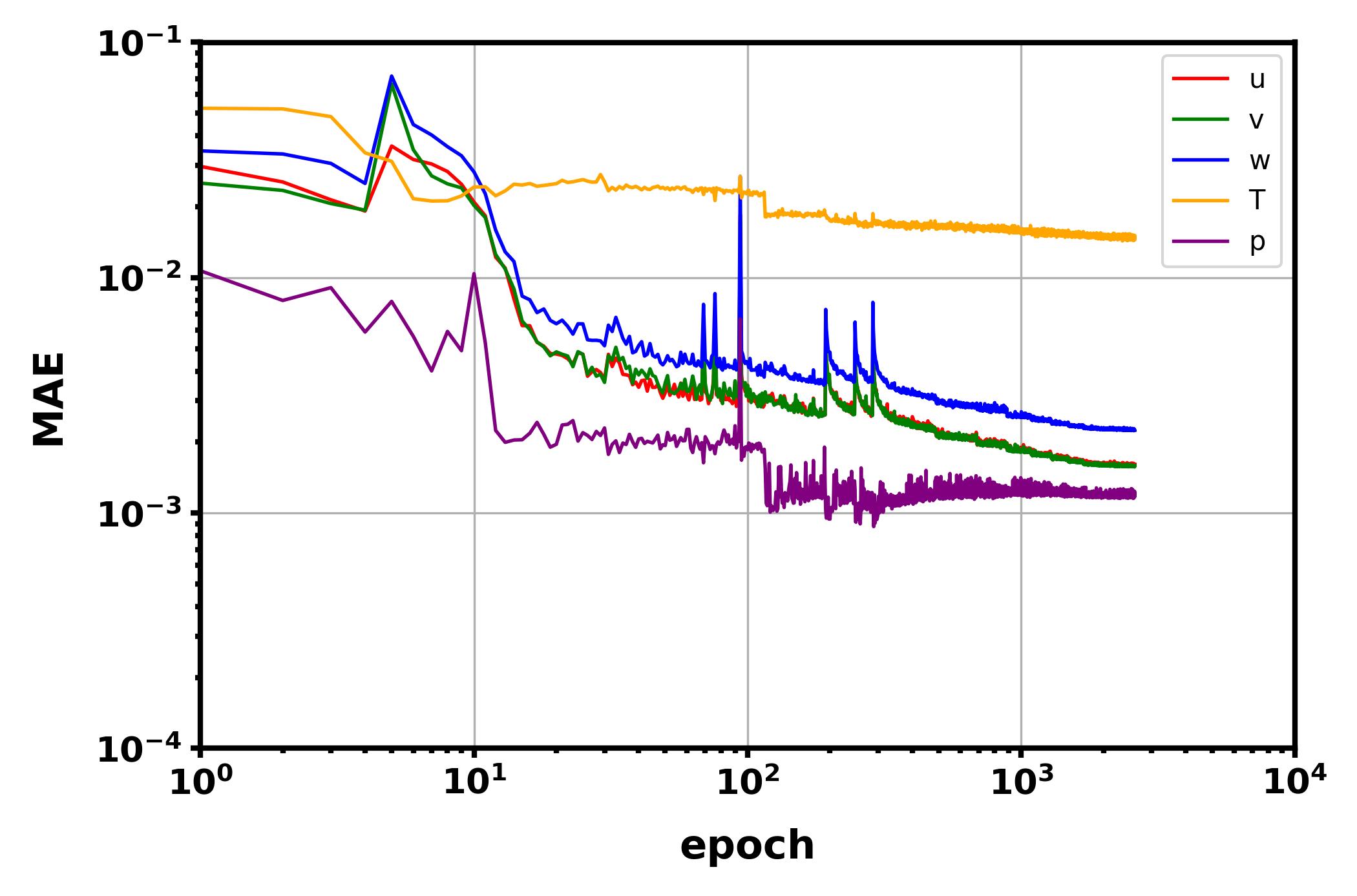}
    \caption[PINN PCC DNS case 2]{Pearson correlation values (left) and MAE (right) of the velocity components, temperature and pressure estimated from the reconstructed PINN fields compared with the DNS fields over 2500 epochs of training.}
    \label{CH6F::cor2}
\end{figure}

The correlation values of the output fields computed in the course of the training are shown in the left panel of figure \ref{CH6F::cor2}. The correlation values of the velocity components are nearly 100\% with respect to the ground truth. It is noteworthy that the reconstructed temperature and pressure fields also reach correlation values of about 90\% with respect to the ground truth at the end of training. The correlation values are not biased by the boundary conditions, since they are estimated at the particle positions, and there are no particles at the boundaries in our dataset. Therefore, correlation values are suitable for comparing the flow structures in the reconstructed fields with those of the ground truth (DNS). Also, the MAE of the considered PINN flow fields compared to the DNS fields are shown in the right panel of figure \ref{CH6F::cor2} for the course of the training. The MAE of the velocity components, pressure, and temperature converge to values of about $2\cdot10^{-3}$, $1\cdot10^{-3}$, and $1\cdot10^{-2}$, respectively, after 2500 training epochs. Both the correlation and MAE values are estimated from the last ten time steps of the DNS dataset.

\begin{figure}[H]
    \hspace{0.9cm} DNS \hspace{5.9cm} PINN \vspace{0.3cm} \\
    \centering
        \includegraphics[width=0.445\textwidth]{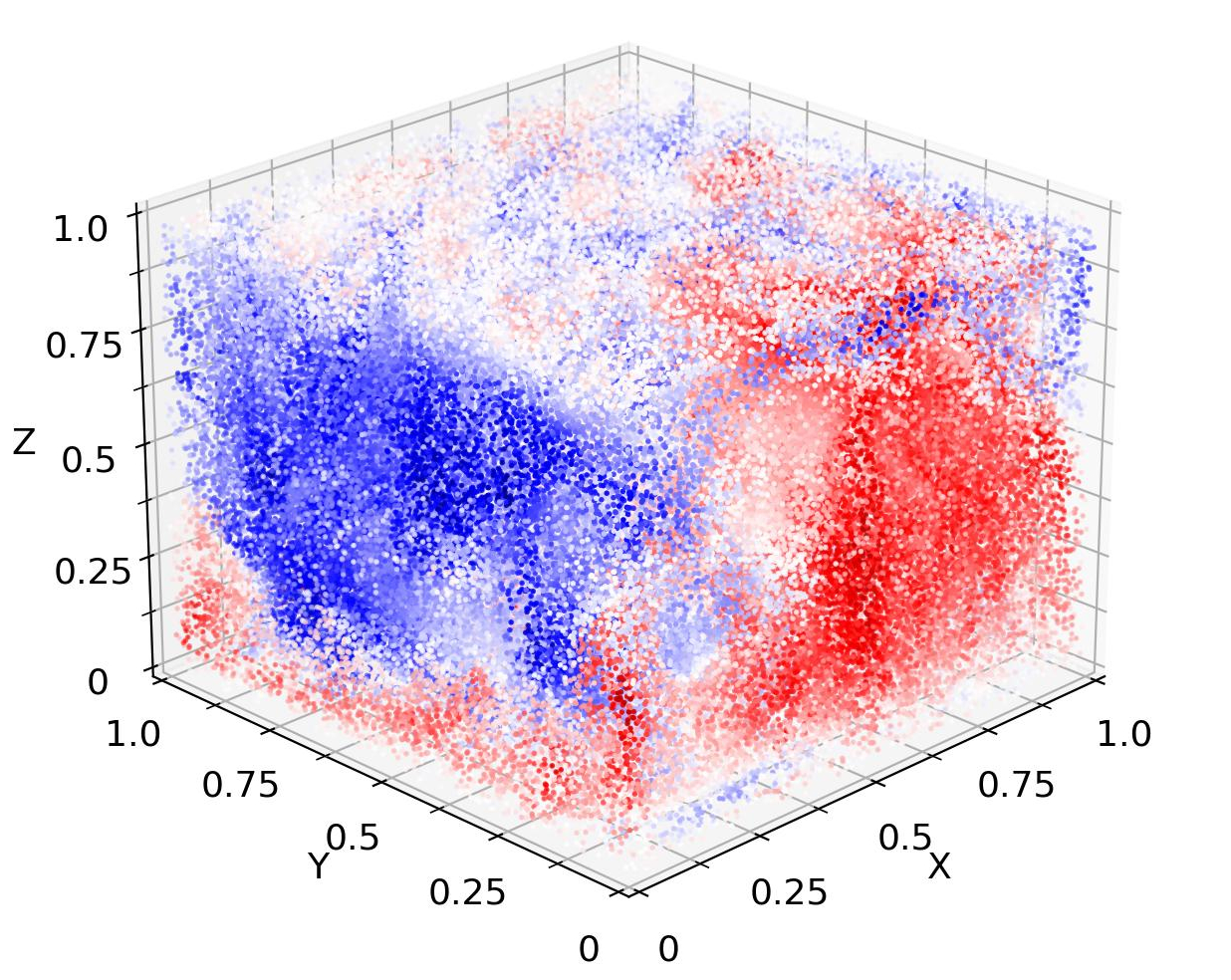}
        \includegraphics[width=0.445\textwidth]{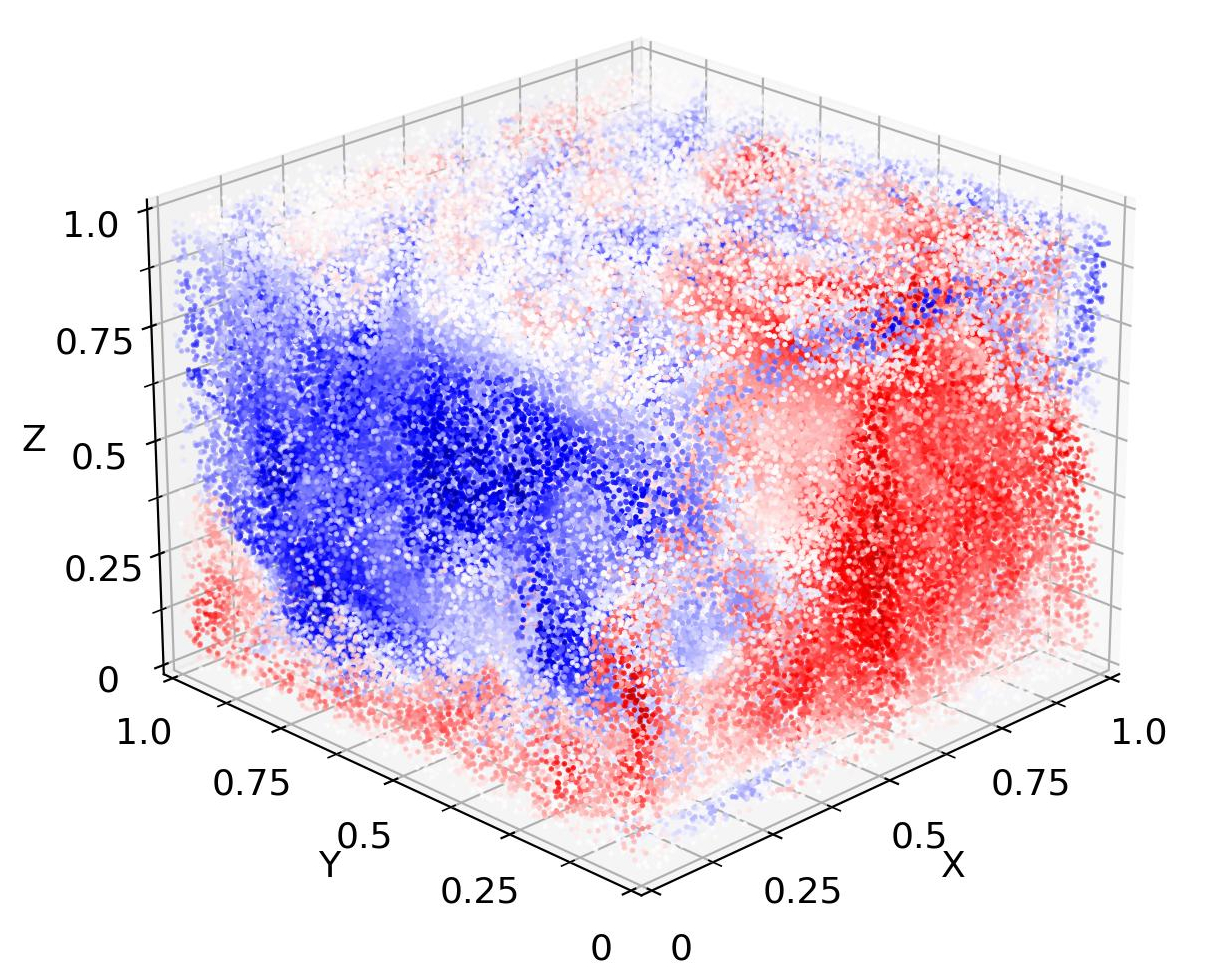} \hspace{0.01cm}
        \includegraphics[width=0.085\textwidth]{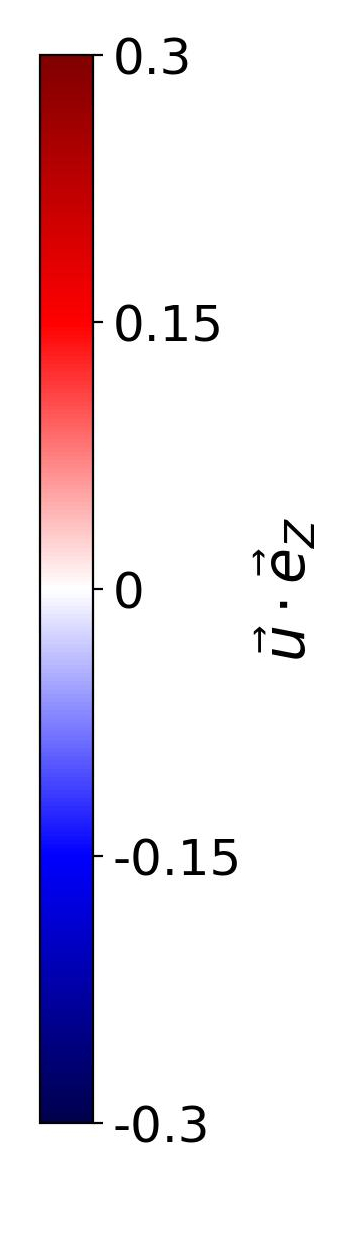}
        \includegraphics[width=0.45\textwidth]{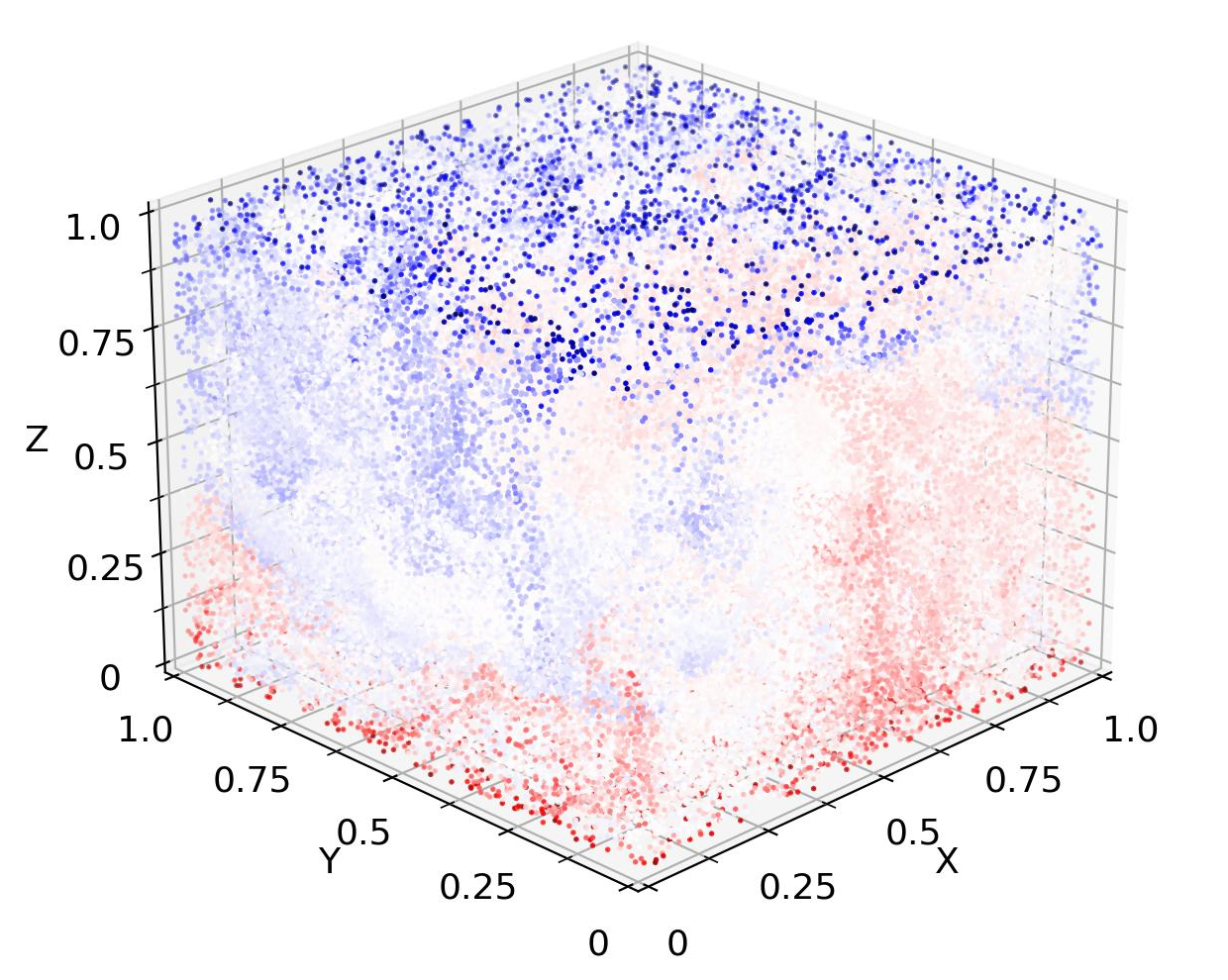}
        \includegraphics[width=0.45\textwidth]{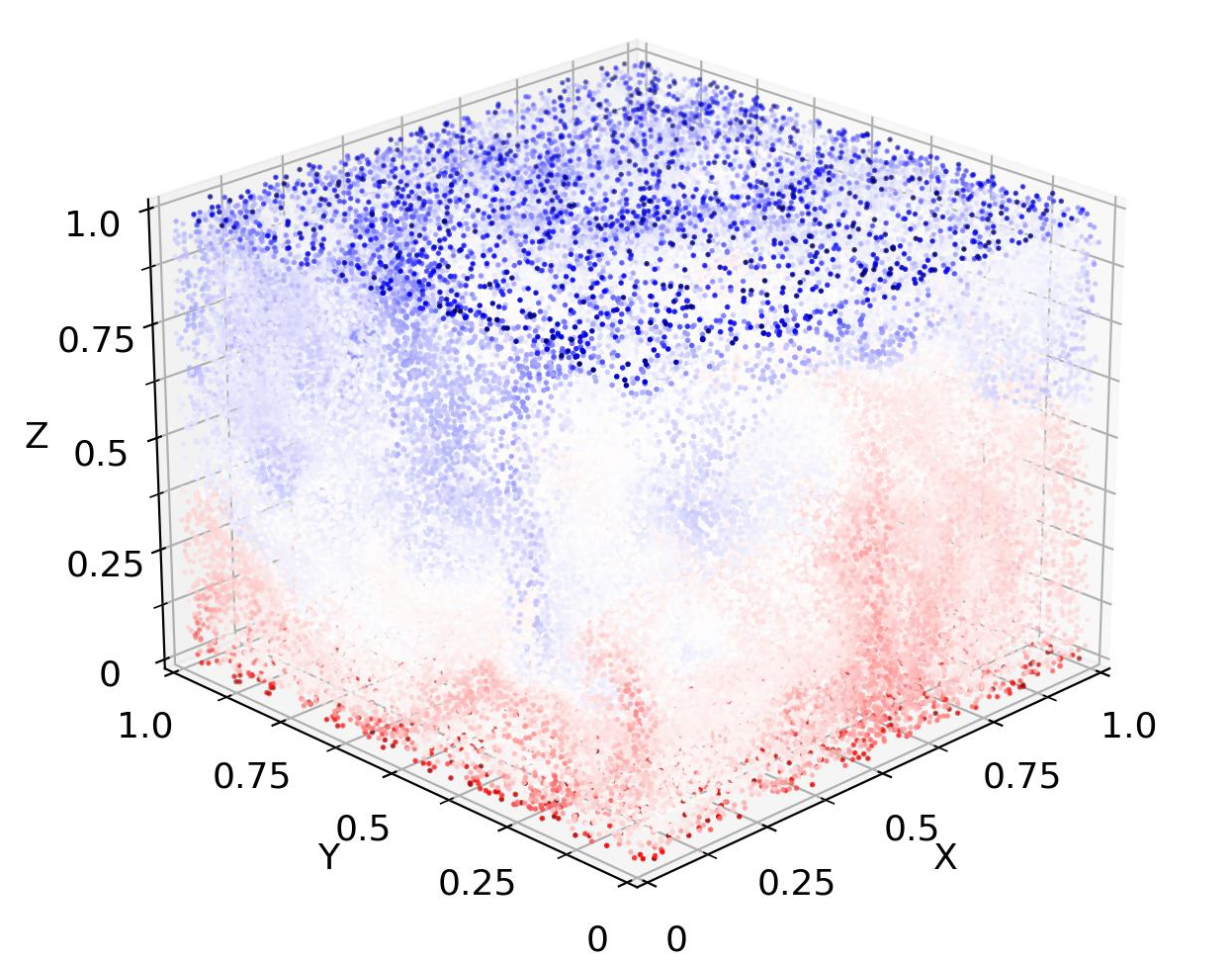}
        \includegraphics[width=0.07\textwidth]{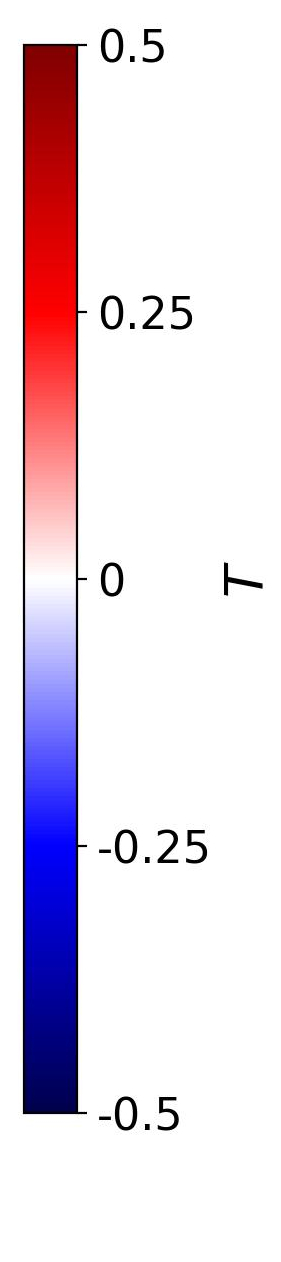}
        \includegraphics[width=0.45\textwidth]{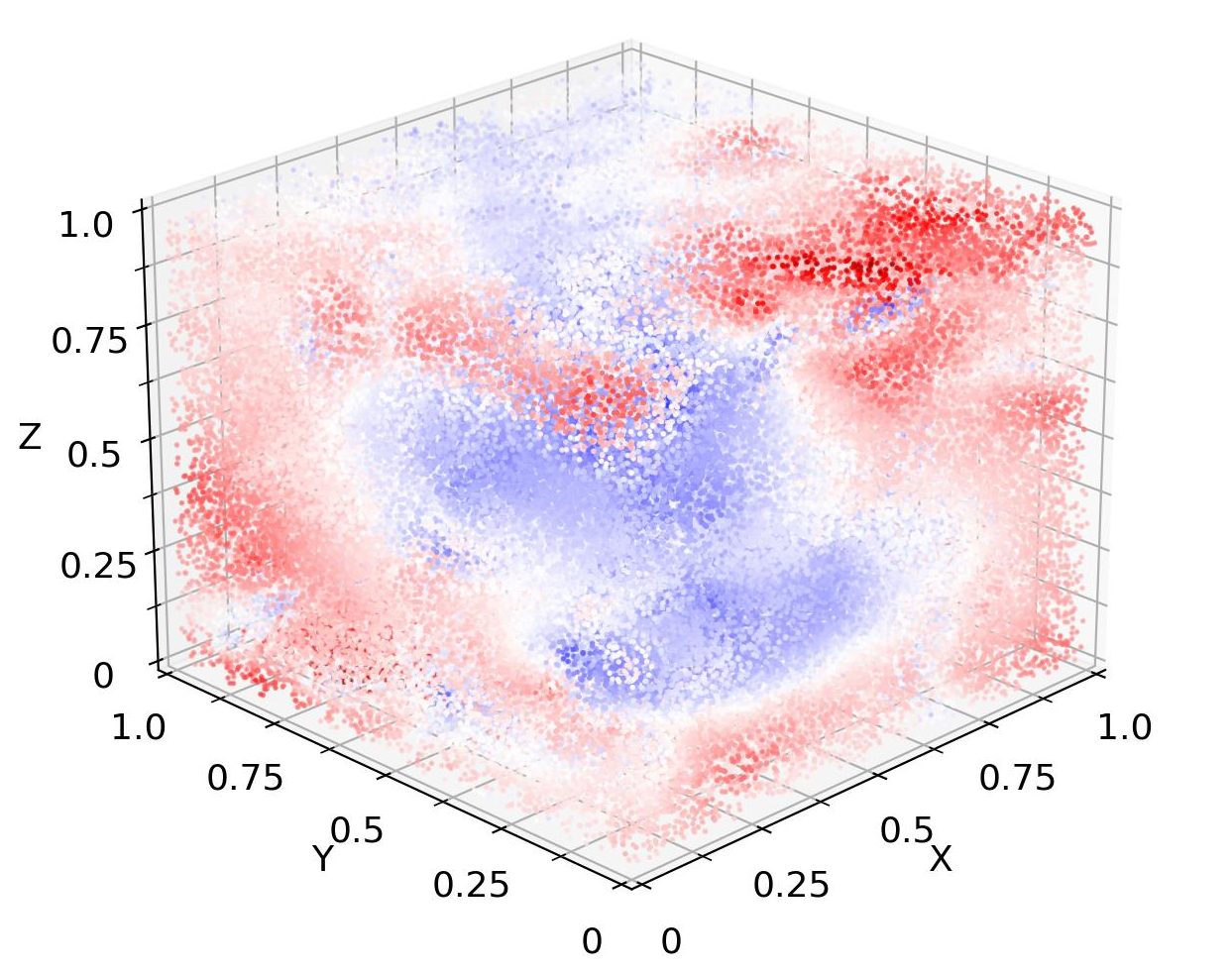}
        \includegraphics[width=0.45\textwidth]{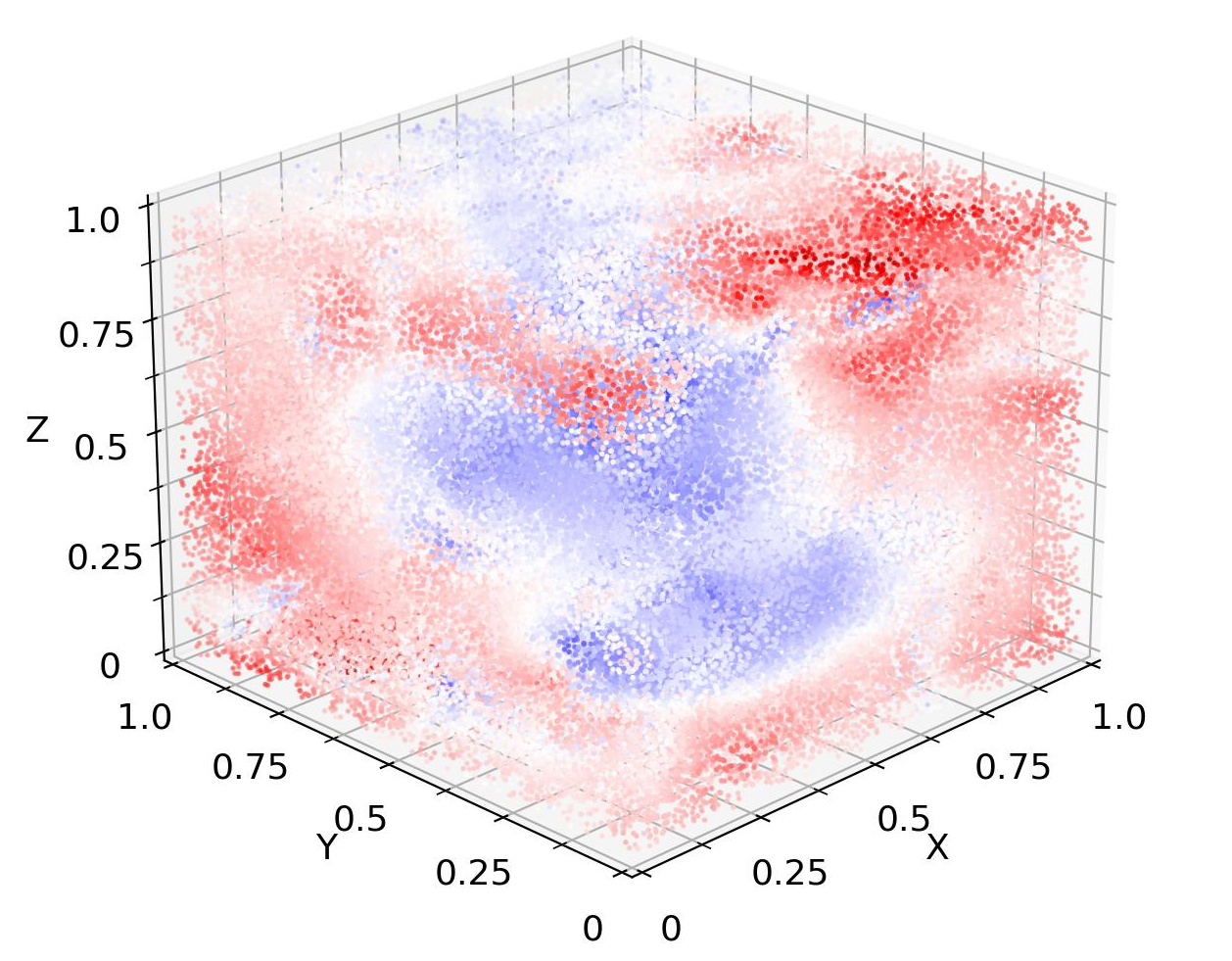}
        \includegraphics[width=0.083\textwidth]{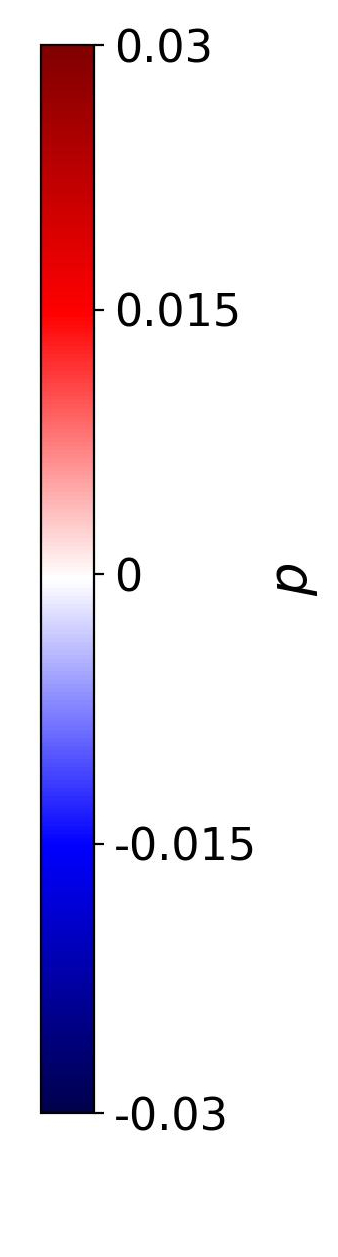}
    \caption[3D compare]{Colored Instantaneous vertical velocity (top), temperature (mid) and pressure (bottom) at the time step 120. Left: DNS vertical velocity, temperature and pressure on the particle positions. Right: reconstructed vertical velocity, temperature and pressure on the particle positions using the PINN.}
    \label{CH6F::compare}
\end{figure}

The high correlation values of the reconstructed properties with respect to the ground truth can be visually investigated in figure \ref{CH6F::compare}, which provides a visual comparison of the instantaneous vertical velocity, temperature, and pressure fields at time step 120 of the DNS and their reconstruction using PINN. It is difficult to see differences in velocity and pressure, by comparing the temperature fields color differences can be seen near the cooling plate and in the smaller plume-like structures in the bulk. 

\begin{figure}[H]
    \centering
        \includegraphics[width=0.9\textwidth]{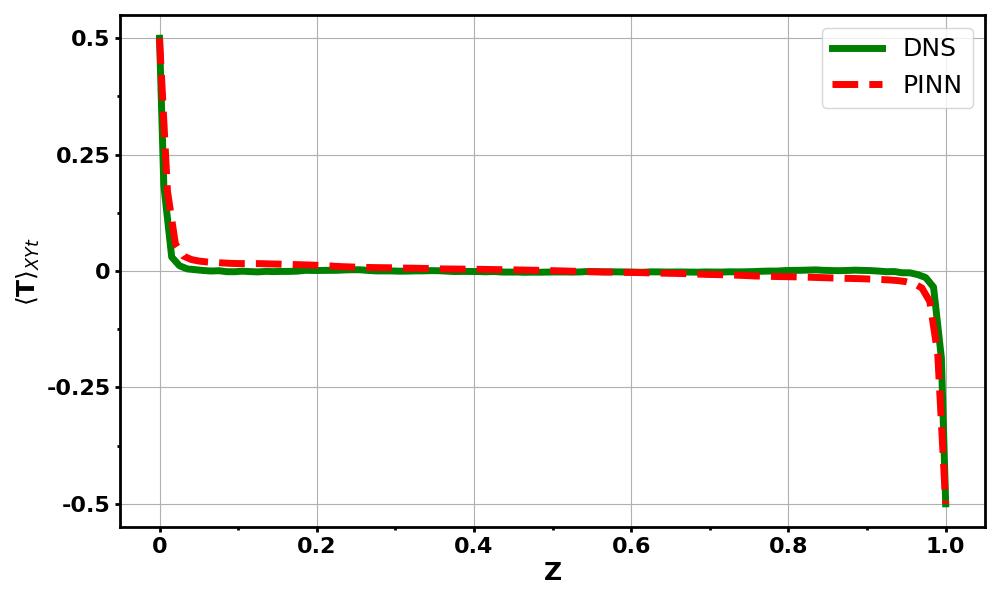}
    \caption[PINN T profile DNS case 2]{Computed and reconstructed mean temperature profiles over the cell height determined from the temperatures of the particles in the DNS and the PINN, respectively.}
    \label{CH6F::Tprof2}
\end{figure}

The differences in the reconstructed temperature field compared to the ground truth temperature are analyzed by comparing their averaged temperature profiles as a function of cell height $Z$, shown in figure \ref{CH6F::Tprof2}. The mean temperature profiles with steep gradients at the walls and the nearly vanishing gradients in the bulk are typical for the considered high Ra number flow with a well mixed bulk region. In principle, the mean temperature profiles agree well, but the profile of the reconstructed temperature deviates from the DNS profile in the boundary layers, as methods to reconstruct the boundary layer as suggested in \cite{volk2025pinn} are not applied. Interestingly, the PINN is trained with the correct mean temperature profile, see figure \ref{CH6F::PINNTmean}, but during training the mean temperature profile deforms slightly due to the added temperature fluctuations in equation (\ref{eq:TmeanPINN}) which are not canceled out entirely in the averaging process over the last 10 time steps of the dataset.

\begin{figure}[H]
    \centering
        \includegraphics[width=\textwidth]{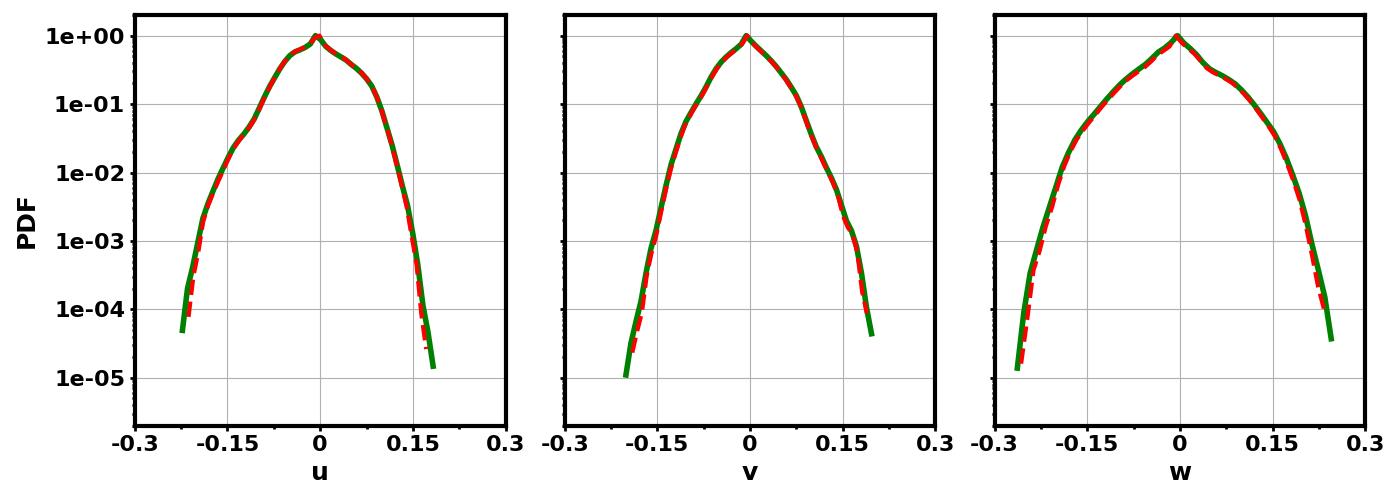}
        \includegraphics[width=\textwidth]{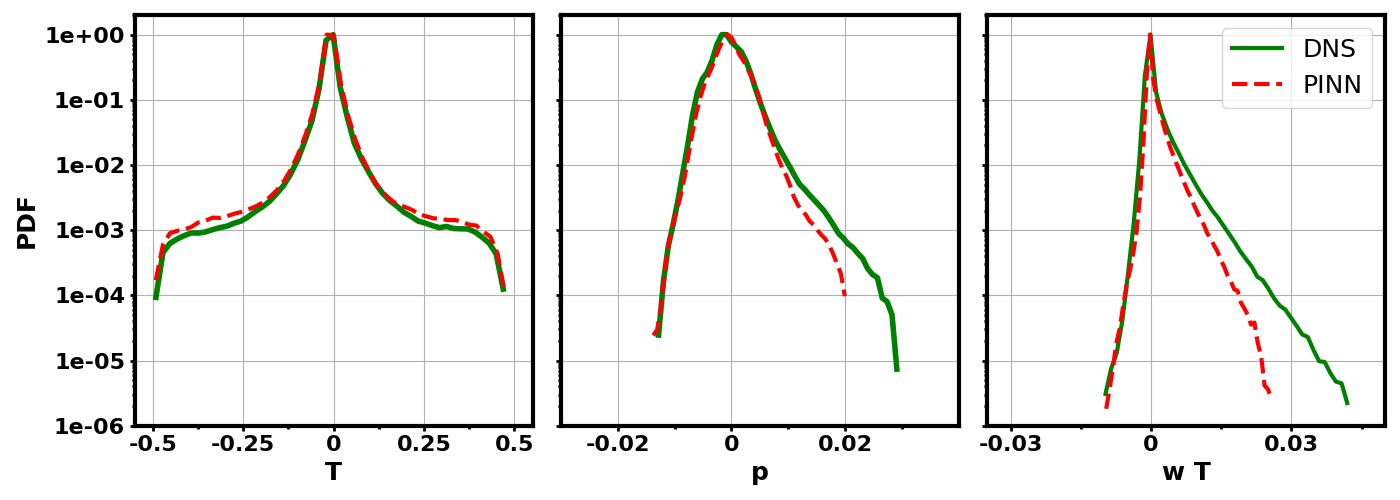}
    \caption[PDF]{PDFs of the three velocity components, temperature, pressure and heat flux calculated from the synthetic PTV tracks from the DNS in green compared with the PDFs obtained with the PINN results in red.}
    \label{fig:PDF}
\end{figure}

Figure \ref{fig:PDF} shows the flow field statistics by comparing the probability density function (PDF) of the reconstructed velocity components, temperature, pressure, and heat flux $w\,T$ using PINN with those of the DNS dataset. In both cases, the last 10 time steps of the dataset are used to generate the data shown. In accordance with the high PCC values at the end of the training, the PDFs of the velocity components obtained from the DNS and the PINN agree well. Also, the PDFs of the temperature shown in figure \ref{fig:PDF} agree well. However, there are some deviations with more frequent large reconstructed temperature values obtained with the PINN than predicted in the DNS. The reason for this is that the PINN was trained with the exact temperature boundaries, which required the PINN to learn a temperature field that connects the bulk with the boundary values. Therefore, deviations from DNS temperature values occur mainly in the boundary layers, where the absolute temperature and its gradient are high. In contrast, the PDF of the reconstructed pressure shows differences for positive pressure values with the DNS pressure, which has a strongly asymmetric distribution around zero, but at zero and negative pressures the PDFs agree well. 

\begin{figure}[H]
    \centering
        \includegraphics[width=\textwidth]{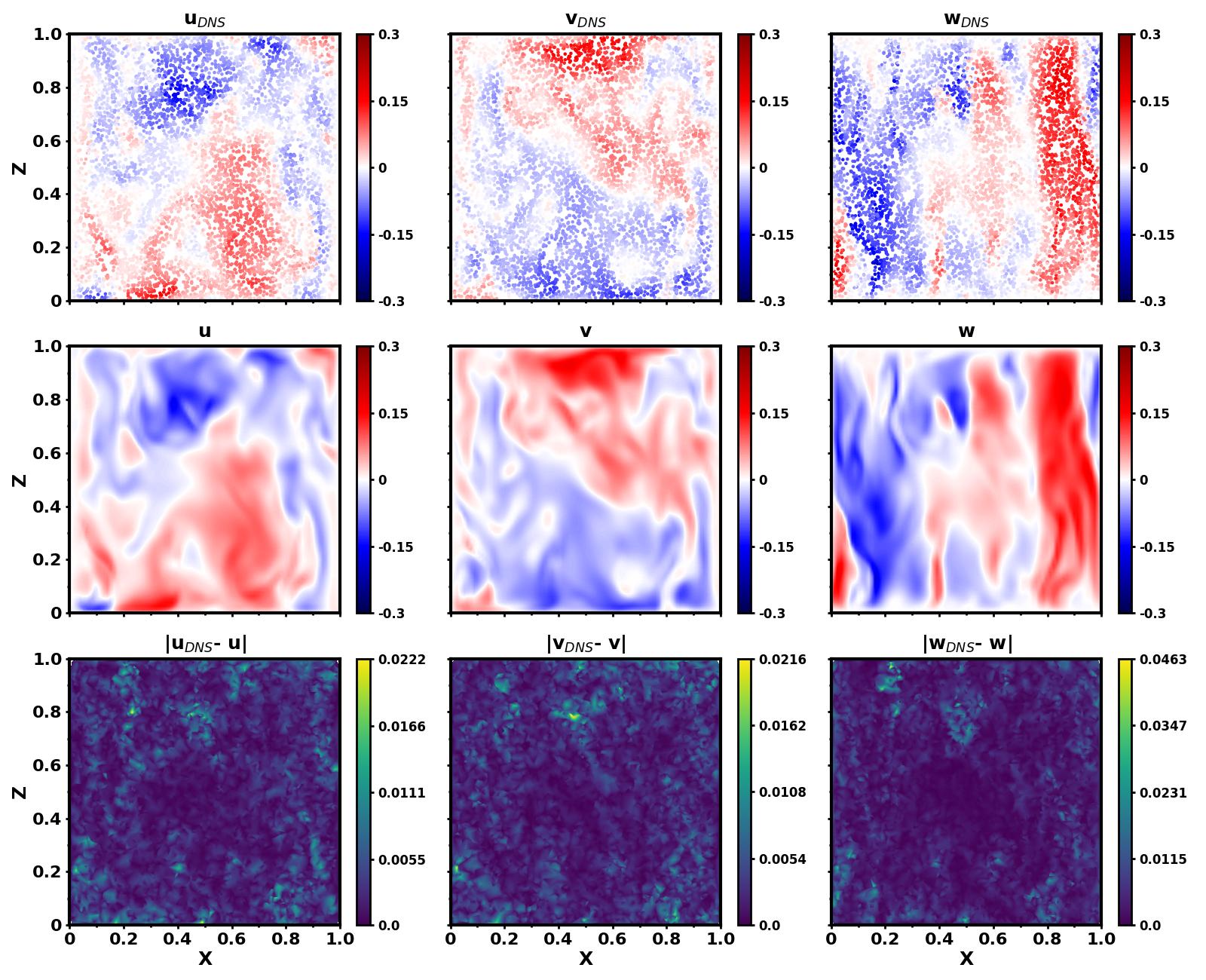}
    \caption[PINN uvw evaluation DNS case 2]{Color contours of the three velocity components in the diagonal plane of the cubic cell containing the LSC with a thickness of $L/50$ predicted in the DNS $u_{DNS},\,v_{DNS},\,w_{DNS}$ (ground truth) in comparison with the reconstructed velocities $u,\,v,\,w$ computed from the synthetic PTV tracks using the PINN. The absolute error between the fields of the same properties is visualized in the bottom row.}
    \label{CH6F::uvw2}
\end{figure}

\begin{figure}[H]
    \centering
        \includegraphics[width=\textwidth]{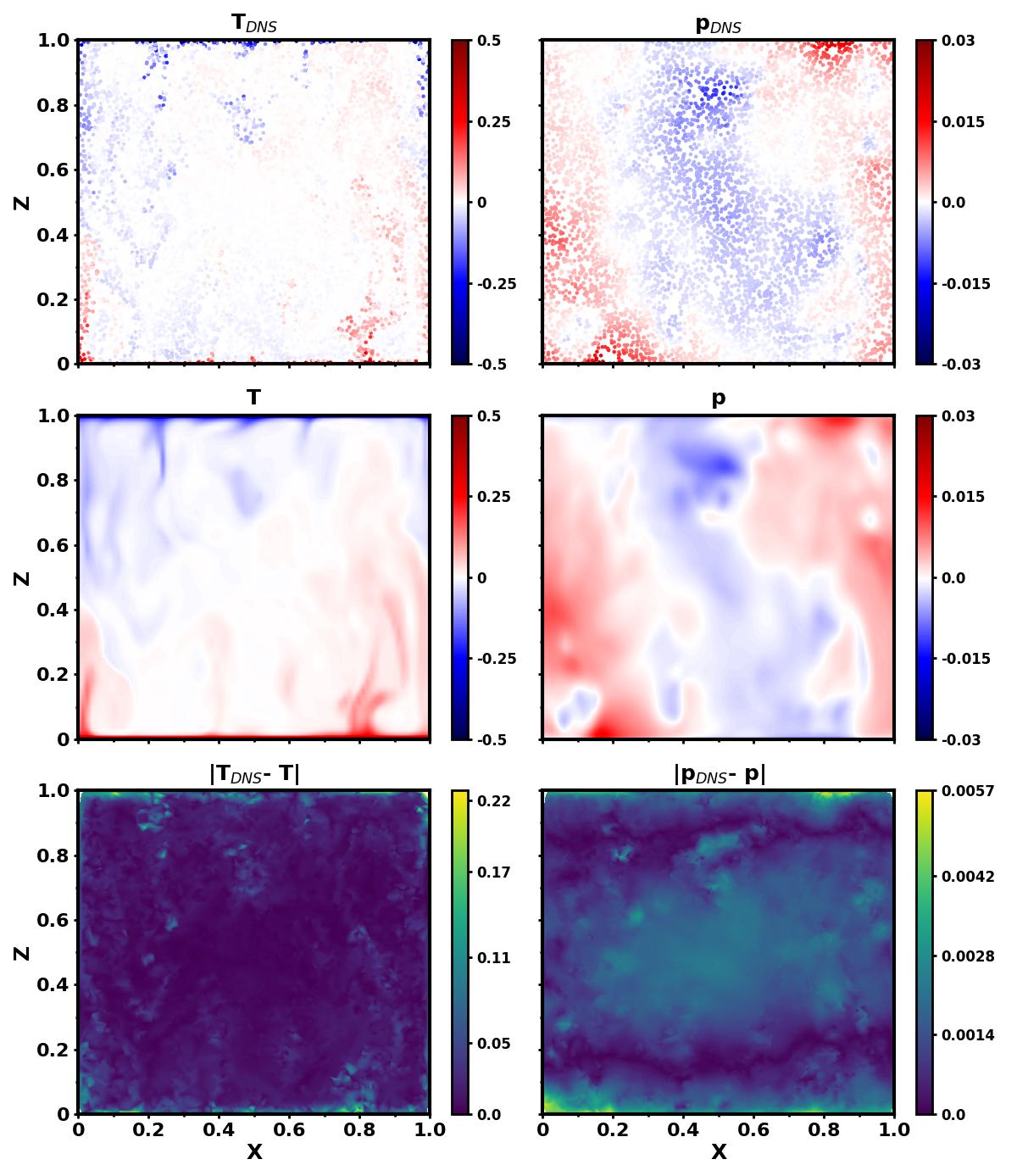}
    \caption[PINN Tp evaluation DNS case 2]{Color contours of the pressure and temperature in the diagonal plane of the cubic cell containing the LSC with a thickness of $L/50$ predicted in the DNS $T_{DNS},p_{DNS}$ (ground truth) in comparison with the reconstructed properties $T,p$ computed from the synthetic PTV tracks using the PINN. The absolute error between the fields of the same properties is visualized in the bottom row.}
    \label{CH6F::Tp2}
\end{figure}

Compared to the overall picture of the flow given in figure \ref{CH6F::compare}, a direct comparison of the reconstructed flow fields with the DNS in the diagonal plane of the cubic cell containing the LSC is given next, since in this plane the most interesting dynamics in the flow can be observed. The corresponding instantaneous DNS velocity components at the particle positions, the reconstructed velocity components, and the difference between these components at time step 120 in the diagonal plane of the cubic cell containing the LSC with a thickness of $L/50$ are shown in figure \ref{CH6F::uvw2} to demonstrate the visual correlation between these fields. Figure \ref{CH6F::Tp2} shows the corresponding temperature and pressure fields and their differences in the same way. As shown in the bottom row of figure \ref{CH6F::uvw2}, the reconstructed velocity components differ only slightly from the ground truth, consistent with the high PCC values of nearly 100\% at the end of training. The error distribution of all velocity components looks similar and only a few patches with a maximum error of about 10\% can be found, overall the average relative error with respect to the maximum velocity of $0.3$ is about 1\% for all velocity components, which is a good result showing that the training of the PINN with respect to the data loss worked perfectly. 

The reconstructed temperature and pressure show the highest error near the heating and cooling plate compared to the ground truth, see the bottom row in figure \ref{CH6F::Tp2}. In the bulk, both fields show small errors and the average relative error is about 5\% for the temperature and 10\% for the pressure. The fact that the largest reconstruction errors are in the boundary layers is consistent with previously discussed results and is due to the lack of data from a well-resolved boundary layer and the lack of a method to correctly reconstruct the boundary layers. Most importantly, the reconstructed temperature and pressure fields behave realistically. The temperature field shows thermal plumes that extend from the heating and cooling plates to the entire cell height, which are also coupled to the motion of the LSC. The pressure field is characterized by pressure minima at the positions of the secondary circulations in the corner and in the center of the LSC. Pressure maxima are correctly reconstructed at the position where the LSC interacts with the boundaries.

The physical response in RBC is measured by the Nusselt number, which describes the dominance of convective heat transport over conductive heat transport, and we know that the time and space averaged Nusselt number in the DNS is approximately $\text{Nu}=63.4$ for system parameters Ra $=10^9$ and Pr $=6.9$. Thus, we can verify whether the PINN reconstructs a flow with the correct physical heat flux over the cell height by calculating the Nusselt number with equation (\ref{eq:nusselt}), resulting in $\text{Nu}_{\text{PINN}}=61.8$, using the last 10 time steps of the dataset for the space and time average. The resulting Nu deviates only a little from the Nu value obtained in the DNS. Thus, we can expect that the PINN provides a realistic flow. Note that the averaging interval is not sufficient for complete statistical convergence of Nu in the current dataset.



\section{Temperature and pressure reconstruction using experimental PTV data}
\label{exp2}
In the following, the temperature and pressure reconstruction with the proposed PINN is tested using the velocity fields of an experimental PTV dataset (Ra $\approx10^9$, Pr $\approx6.9$), see section \ref{exp}, consisting of nearly 300000 particle trajectories over 150 time steps, of which about 50000 are active per time step. The dataset covers a similar time span as the DNS dataset of 2.5 free fall time units. The parameter $a=126$ is chosen to mimic the mean temperature profile and temperature boundary conditions at the top and bottom plates for the considered case with a Nusselt number Nu $\approx 63$, known from the DNS for similar system parameters. The PINN was trained over 2500 epochs and the training process took about 38 seconds per epoch on a NVIDIA GeForce RTX 4090. 

\begin{figure}[H]
    \centering
        \includegraphics[width=0.9\textwidth]{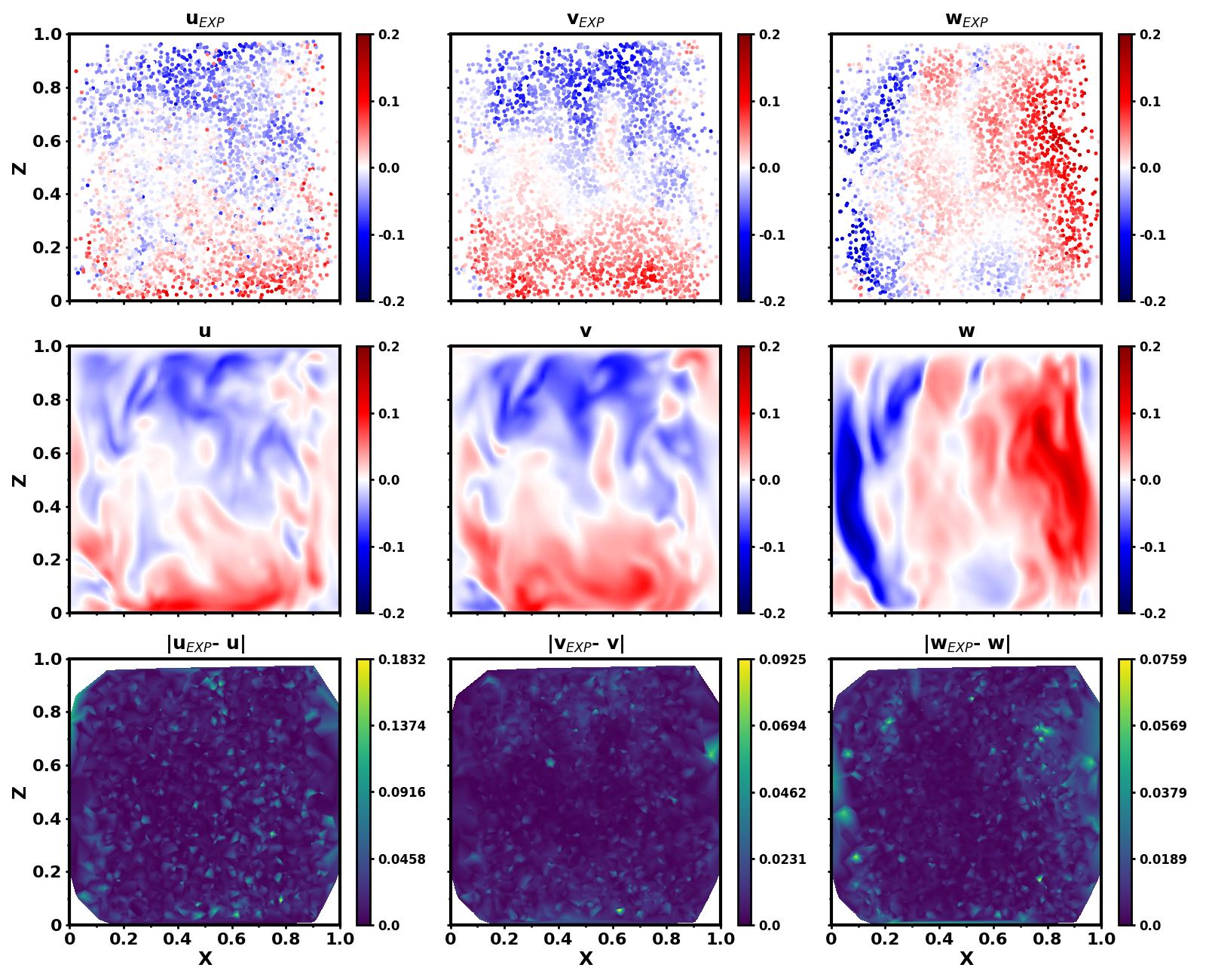}
    \caption[PINN uvw evaluation EXP case 2]{Color contours of the three velocity components in the diagonal plane of the cubic cell containing the LSC with a thickness of $L/20$ measured with PTV experimentally $u_{EXP},\,v_{EXP},\,w_{EXP}$ in comparison with the reconstructed velocities $u,\,v,\,w$ computed using the PINN. The absolute error between the fields of the same properties is visualized in the bottom row.}
    \label{CH6F::uvw}
\end{figure}

\begin{figure}[H]
    \centering
        \includegraphics[width=0.9\textwidth]{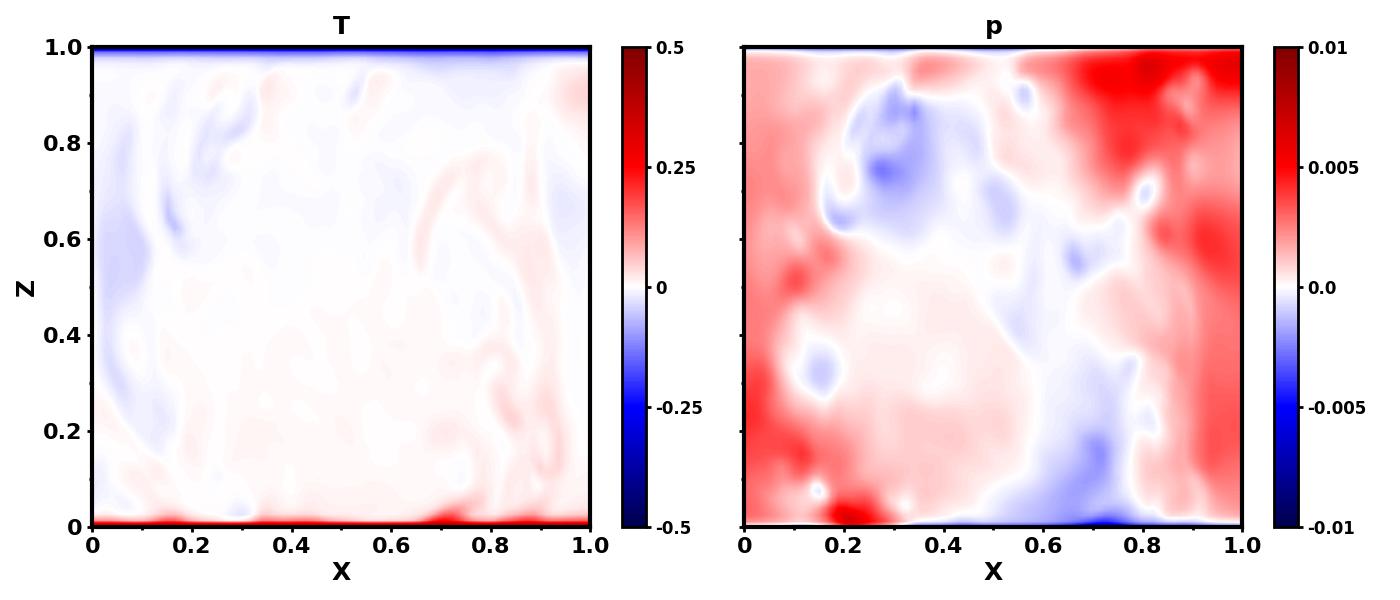}
    \caption[PINN Tp evaluation DNS case 2]{Temperature field $T$ and pressure field $p$ in the diagonal plane of the cubic cell containing the LSC constructed with the PINN based on experimental PTV tracks.}
    \label{CH6F::Tp}
\end{figure}

The particle positions and velocities measured in the PTV experiments are associated with calibration and image processing errors. It is well known that the velocity component $u$ in the direction of the cameras is associated with the largest error. Thus, reconstructing physical flow fields with a PINN provides a method for processing experimental data while maintaining physical consistency with the governing equations (\ref{eq:momentum})-(\ref{eq:mass}) and reducing measurement uncertainties, as shown in figure \ref{CH6F::uvw} where the measured velocities (top row) are compared with the reconstructed velocity (middle row) at the particle positions in the diagonal plane of the cubic cell containing the LSC with a thickness of $L/20$ at time step 145. The bottom row shows the absolute error of the reconstructed velocities with respect to the measured velocities. It is clear that the experiment does not resolve boundary layers, since the particle density reduces to zero near the sidewalls in figure \ref{CH6F::uvw}. However, the PINN can reconstruct unmeasured flow properties. In our case this means to construct the temperature and pressure fields, which are shown in figure \ref{CH6F::Tp} in the diagonal plane of the cubic cell containing the LSC at time step 145. The reconstructed temperature and pressure fields show realistic results compared to figure \ref{CH6F::Tp2}. The temperature field features thermal plumes of integral length that detach from the heating and cooling plates and are coupled to the motion of the LSC. In addition, pressure minima are located in the center of the LSC, and in corner circulations, while pressure maxima are constructed in regions where the LSC interacts with the boundaries. 

\begin{figure}[H]
    \centering
        \includegraphics[width=\textwidth]{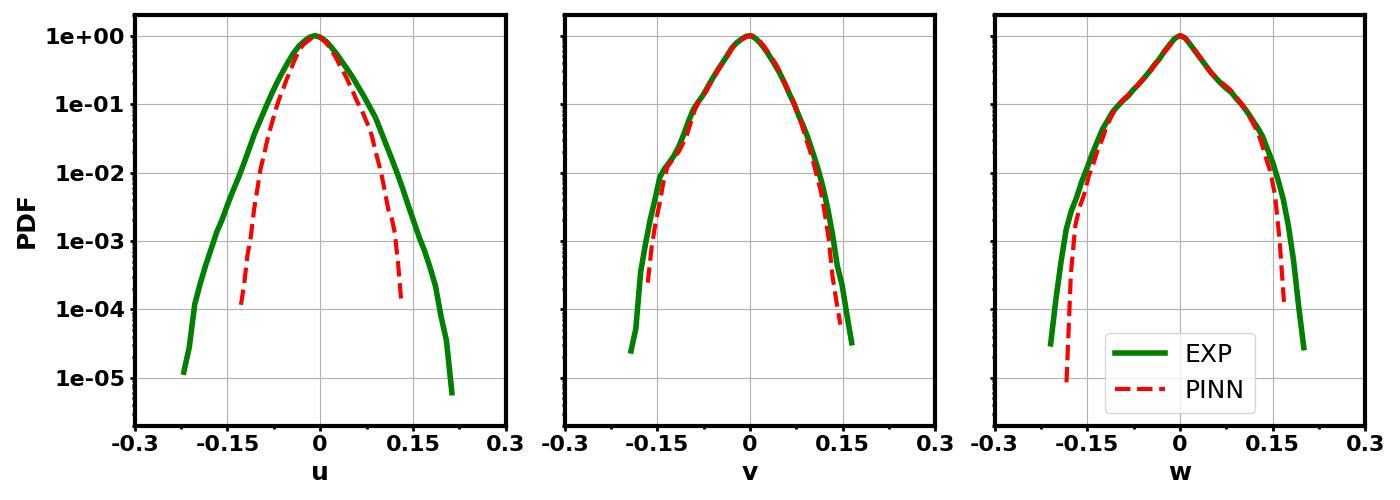}
        \includegraphics[width=\textwidth]{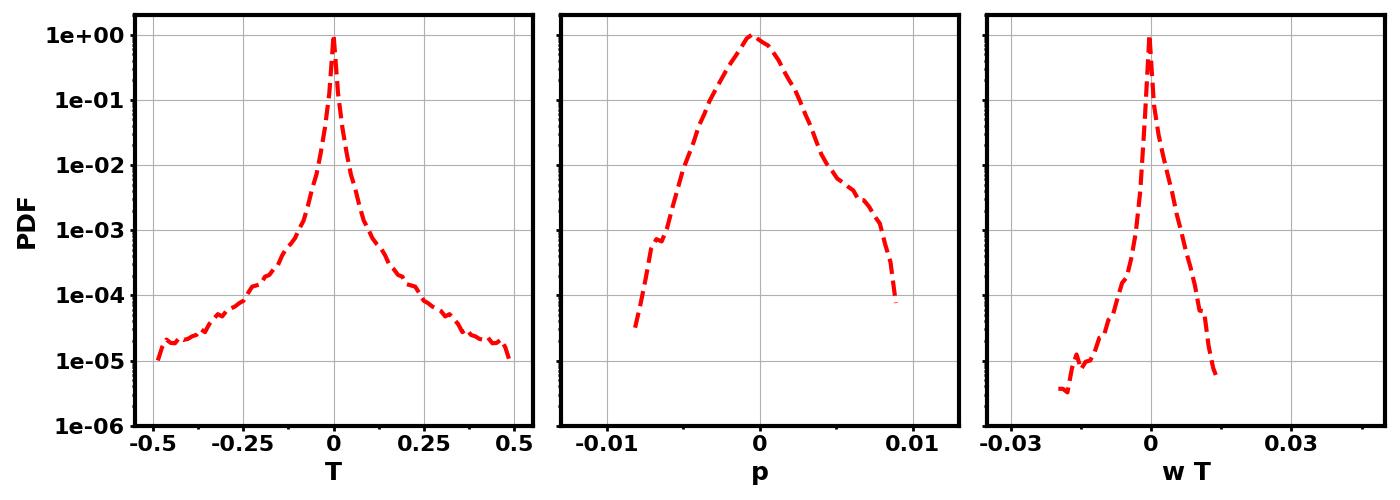}
    \caption[PDFexp]{PDFs of the three velocity components, temperature, pressure and heat flux calculated from the experimental PTV tracks in green compared with the PDFs obtained with the PINN results in red.}
    \label{fig:PDFexp}
\end{figure}

Figure \ref{fig:PDFexp} shows the probability density function (PDF) of the constructed velocity components, temperature, pressure, and heat flux $w\,T$ using the PINN. In the top row of figure \ref{fig:PDFexp} the PDFs of the velocity components are compared  with those obtained from the measured velocities. The PINN reconstructs a similar distribution for the velocity components $v$ and $w$ as provided by the experiment. 

The reconstructed distribution of the $u$ component shows less extreme values, as they do not represent the correct physics, and are mainly due to measurement errors in the direction of the cameras. Thus, the PINN is capable of denoising experimental data while maintaining physical consistency. Since the $u$ component has measurement uncertainties, it can help to fine-tune the corresponding data loss weight in the loss term during training, separate from the lambda values of the other velocity components. However, no significant improvements are found in a direct test, and a thorough investigation of the weight factor choice may be helpful. The bottom row of figure \ref{fig:PDFexp} shows the PDFs of the not directly measured temperature, pressure, and convective heat flux. The PDF of temperature has a characteristic bimodal distribution that does not have as many high absolute temperature values as the PDF of temperature in the DNS, see figure \ref{fig:PDF}. This discrepancy can be attributed to the absence of boundary layers in the experimental data, which the PINN consequently did not reconstruct. The highest absolute temperature values are typically concentrated within these boundary layers, which is explains their absence in the reconstructed temperature distribution. The pressure PDF is similar to the DNS pressure PDF, and the heat flux PDF has the same shape as the DNS heat flux PDF, but like the temperature PDF, with far fewer high absolute values.

The physical response of the reconstructed flow is evaluated using the reconstructed vertical velocity and temperature to calculate the averaged Nusselt number using equation (\ref{eq:nusselt}), resulting in $\text{Nu}_{\text{PINN}}=34.6$, using the last 10 time steps of the dataset for the space and time average. Note that the averaging interval is not sufficient for complete statistical convergence of Nu in the current dataset. The deviation from the reference value at Nu $\approx63$ is explained by examining the color scales of the particle tracks in figures \ref{fig:tracks} and \ref{fig:tracksdns}, where it can be seen that the vertical velocity scale of the DNS tracks is about a factor of 1.6 larger than that of the measured tracks. In an ideal setup, such as the DNS, RBC involves adiabatic side walls, which are absent in our experiment. Instead, we used glass sidewalls with 8 mm thickness to provide optical access for PTV measurements. This configuration results in an estimated total relative heat loss through the side walls of approximately 20\%. This heat loss diminishes a portion of the buoyancy forcing that drives the LSC, which consequently reduces the maximum normalized vertical velocity compared to DNS. This reduction is clearly observable: the maximum normalized vertical velocity is about 0.29 in the DNS versus 0.18 in the experiment. Applying the corrective factor of 1.6 to the estimated averaged Nusselt number yields Nu $=55.3$. The remaining discrepancy between this value and the expected value of about 63 is based on measurement uncertainties in the experimental procedure and reconstruction errors of the proposed method.

\section{Conclusion}
\label{concl}
This study successfully demonstrates the application of a physics-informed neural network (PINN) to reconstruct temperature and pressure fields from Lagrangian velocity data in turbulent Rayleigh-B\'{e}nard convection (RBC) at Ra = $10^9$ and Pr = 7, utilizing both synthetic direct numerical simulation (DNS) and experimental particle tracking velocimetry (PTV) datasets. The PINN, employing a multilayer perceptron architecture with a periodic sine activation function, incorporates a modification by adding a mean temperature profile to the temperature output, as defined in equation (\ref{eq:TmeanPINN}). This profile, parameterized by $a=2\,\text{Nu}$, ensures exact satisfaction of the temperature boundary conditions and facilitates accurate reconstruction of temperature fluctuations, particularly in the well-mixed bulk region, thus addressing the numerical challenges posed by high temperature gradients in the hard turbulence regime.

In the DNS case, the PINN is able to reconstructed temperature and pressure fields with a correlation of 90\% with respect to the ground truth. For this high correlation factor the associated flow fields consist of thermal plumes, large-scale circulation, and pressure variations at the correct locations as the comparison with DNS data shows. However, the reconstruction effectiveness is limited in the thermal and viscous boundary layers due to insufficient particle data near the walls and the PINN's lack of specialized training for these regions. This limitation results in deviations in the temperature profiles near the boundaries, as observed in figure \ref{CH6F::Tprof2}. When applied to experimental PTV data, the PINN not only constructs physically consistent temperature and pressure fields but also reduces measurement uncertainties in the velocity fields, demonstrating its robustness in handling noisy experimental data. The reconstructed Nusselt number profile in the bulk flow deviates from the expected value due to heat losses through non-adiabatic glass sidewalls, which diminish buoyancy-driven flow compared to the ideal DNS setup. However, PINN reproduces realistic integral flow structures, such as thermal plumes and pressure minima in circulatory regions, highlighting its ability to assimilate unmeasured flow properties in complex convective systems. The compatibility of the PINN with the open-source proPTV framework represents a significant advancement, enabling comprehensive flow analysis without direct measurements of all flow quantities. The PINN code, available at \url{https://github.com/DLR-AS-BOA}, supports further development and application across diverse flow scenarios. While the proposed methodology excels in reconstructing bulk flow dynamics, future improvements should focus on enhancing boundary layer reconstruction, potentially by adopting and refining the boundary-focused training strategies proposed by \cite{volk2025pinn} or incorporating additional physical constraints to account for experimental non-ideal, such as non-adiabatic boundaries. These advancements will further strengthen the applicability of PINNs in studying turbulent thermal convection and other complex flow systems.

\bibliographystyle{plainnat}
\bibliography{refs}
\end{document}